\documentclass[fleqn,useAMS,usenatbib]{mnras}

\usepackage{newtxtext}

\usepackage[T1]{fontenc}
\usepackage{ae,aecompl}

\usepackage{color}
\usepackage[english]{babel}
\usepackage{graphicx}
\usepackage{amsmath}
\usepackage{amssymb}
\usepackage{bm}
\usepackage[utf8]{inputenc}
\usepackage{ulem}

\begin{document}

\newcommand{\dd}{\mathrm{d}}
\newcommand{\rhob}{\rho_\text{b}}
\newcommand{\Tb}{T_\text{b}}
\newcommand{\kB}{\mbox{$k_\text{B}$}}
\newcommand{\rate}{\mbox{erg cm$^{-3}$ s$^{-1}$}}
\newcommand{\gcc}{\mbox{g~cm$^{-3}$}}
\newcommand{\xr}{x_\text{r}}
\newcommand{\gs}{g_\text{s}}
\newcommand{\me}{m_\text{e}}
\newcommand{\msun}{\mbox{$\textrm{M}_\odot$}}


\title[Heat diffusion in neutron star crust]{Model of heat diffusion in the outer crust of bursting
neutron stars}

\author[D. G. Yakovlev et al.]{{D. G.    Yakovlev$^{1}$\thanks{E-mail:
yak.astro@mail.ioffe.ru}},
{  A. D. Kaminker$^1$}, A. Y. Potekhin$^1$,  
{P. Haensel$^{2}$}\\
$^{1}$ Ioffe Institute, Politekhnicheskaya 26, St~Petersburg 194021,
Russia\\
$^{2}$ Copernicus Astronomical Center, Bartycka 18, 00-716 Warsaw, Poland\\
}

\date{Accepted . Received ; in original form}
\pagerange{\pageref{firstpage}--\pageref{lastpage}} \pubyear{2010}

\maketitle \label{firstpage}

\begin{abstract} 

We study heat diffusion after an energy release in a deep spherical
layer of the outer neutron star crust ($10^7 \lesssim \rho \lesssim 4
\times 10^{11}$ \gcc). We demonstrate that this layer possesses specific
heat-accumulating properties, absorbing heat and directing it mostly inside
the star.  It can absorb up to  
$\sim 10^{43}-10^{44}$~erg
due to its high heat
capacity, until its temperature 
exceeds $T \sim 3 \times 10^9$ K
and triggers a rapid neutrino  cooling. A warm layer 
($T \sim 10^8-3 \times 10^9$ K) 
can serve as a good heat reservoir, which is thermally decoupled from
the inner crust and the stellar core for a few months. We present a toy
model  to explore the heat diffusion within the heat-accumulating layer, 
and we test this model using numerical simulations. We formulate some
generic features of the heat propagation which can be useful, for
instance, for the interpretation of superbursts  in accreting neutron
stars. We present a self-similar analysis of late afterglow after such
superbursts, which can be helpful to estimate properties of bursting
stars.     

\end{abstract}



\section{Introduction}
\label{s:introduc}

Many neutron stars
demonstrate bursting activity. For instance, accreting
neutron stars in low-mass X-ray binaries 
show X-ray bursts and superbursts powered
by explosive burning of accreted hydrogen and helium in surface 
layers and subsequent more 
powerful burning of carbon in 
deeper layers (e.g., \citealt{2017Zand,2017Galloway}). These 
processes involve complicated physics of thermal evolution of
accreting neutron stars, steady-state and explosive nuclear 
burning with extended reaction networks, various mass and
heat transport mechanisms (hydrodynamical motions, convection, thermal
diffusion) and
so on. 

We mainly focus on heat diffusion after energy generation in deep layers
of the outer crust of neutron stars. Such a  process has been 
extensively simulated numerically and semi-analytically for about two
decades in the context of modeling superbursts; see, e.g.
\citet{2004CM,2006Cumming,2011KH,2012Altamirano}; \citet*{2012Keek};
\citet{2015Keek} and references therein.  

The outer crust 
(e.g. \citealt*{HPY2007}) is a relatively thin layer which extends from the
stellar surface
to the neutron drip density ($\rho_\text{drip} \approx 4.3 \times 10^{11}$
g~cm$^{-3}$). Its width is only some hundred meters, and
its mass is $\sim 10^{-5}$ \msun. It consists of electrons
and ions (atomic nuclei). We call it crust for simplicity; actually,
the atomic nuclei can constitute either Coulomb solid,
or Coulomb liquid or gas, depending on density $\rho$, temperature $T$
and nuclear composition (our `crust' includes thus the liquid `ocean').
We consider a spherically symmetric star,
neglecting the effects of magnetic fields and rotation.
We will mainly study heat propagation at 
\begin{equation}
\rhob \lesssim \rho \lesssim \rho_\text{drip},~~
   10^8 \lesssim T \lesssim 3 \times 10^9~{\rm K},
\label{e:rho,T}
\end{equation}
where {  $\rhob \sim 10^7$ \gcc\ 
(so that the electrons are relativistic and strongly degenerate)},
and the ions are fully ionized.
Lower $T$ are less interesting as far as
the processes of energy release are concerned (typical ignition 
temperatures for deep nuclear explosions are not so low). We will
analyse specific
heat-accumulating properties of these layers.     

In our previous studies
(e.g. \citealt{2014KAM}; \citealt*{2018Chaikin}, and references therein) 
we have simulated the 
heat propagation in a neutron star after some energy
release in its crust (in 1D and 2D geometries, with 
the heater placed  within a spherical layer or some spot-like region). 
There we have mainly considered the 
heaters that operate quasi-statically over months or longer, 
corresponding either to hypothetical energy release in the crust of
magnetars or to the outbursts (accretion periods) in soft X-ray
transients.

Here we study the heaters that are active 
on time-scales of minutes that is closer to the individual
X-ray bursts or superbursts on neutron stars. {  Our aim is 
to present a simplified model of heat diffusion and test it using a
modern thermal evolution code. The model reproduces and elucidates 
generic properties of deep superbursts and enables one to estimate
how these properties depend on system parameters, particularly
on neutron star mass and radius.}

In Sections \ref{s:toy} and \ref{s:thin-instant} we formulate a simplified
 model for
studying heat diffusion in the $\rho-T$ domain (\ref{e:rho,T}) and discuss its
formal semi-analytic solution for an instant burst in a thin layer. 
Section \ref{s:burstAA} is devoted to
bursting layers of finite width in  domain (\ref{e:rho,T}), 
and Section \ref{s:thick} to bursts in similar layers but extended to lower
densities. We compare analytic solutions with numerical models.
In Section \ref{s:generic} we discuss generic features of heat
diffusion after bursts. In Section \ref{s:fading} we 
analyse late burst decay and present a simple method for evaluating
parameters of bursting neutron stars from observations of such decays. We conclude 
in Section \ref{s:conclude}, and
present some technical details in Appendix~\ref{s:green}.

\section{Simplified analytic model of heat diffusion}
\label{s:toy}

\subsection{Basic parameters and microphysics}
\label{s:toymicro}

We introduce a simplified `toy' model of a spherically  symmetric outer crust
of the neutron star in the $\rho - T$ domain (\ref{e:rho,T}). The crust
is thin and can be regarded as locally flat (e.g. \citealt{1983GPE}).
Unless the contrary is indicated, we will  use this locally flat
coordinate system. Let $z$ be a proper depth measured from the neutron
star surface ($z=0$). The density $\rho$ can be  conveniently expressed
through the relativity parameter of degenerate electrons
\citep{Salpeter61}, $\xr=p_\text{F}/(m_\text{e}c) \approx 1.0088 (\rho_6
Z/A)^{1/3}$, where $\rho_6=\rho/(10^6~\gcc)$, $p_\text{F}$ is the
electron Fermi momentum; $A$ and $Z$ are  the mean ion mass and charge
numbers, respectively. The $A/Z$ ratio is assumed to be constant
throughout the outer crust. With these assumptions, the density profile
$\rho(z)$ is determined by \citep*{HPY2007}
\begin{equation}
\xr^3=\left[ \frac{z}{z_0}\, \left(2 
+ \frac{z}{z_0}   \right) \right]^{3/2}
\approx \left( \frac{z}{z_0} \right)^3,
\label{e:rho(z)}
\end{equation} 
with $z_0=(Zm_\text{e}c^2)/(m_\text{u}\gs A)$ and
$\gs=(GM/ R^2) (1-r_\text{g}/R)^{-1/2}$ ({  $m_\text{u}$
being the atomic mass unit}). Here, $\gs$ 
is the local gravitational acceleration in the
outer crust, which is nearly constant there 
and is expressed through the gravitational mass
of the star $M$ and its circumferential radius $R$;
$z_0$ is a characteristic depth of the outermost layer ($\rho 
\lesssim 10^6$ \gcc) in which the degenerate electrons
are non-relativistic; $r_\text{g}=2GM/c^2$ is the 
gravitational radius of the star.  
The last expression in equation
(\ref{e:rho(z)}) is the asymptote at depths $z \gg z_0$, 
where the electrons are degenerate and ultra-relativistic; 
it will be used below in the toy-model analysis. 
In particular, it gives the column density $y=z \rho/4$; 
it becomes inaccurate at $\rho \lesssim \rhob$.

The diffusion of heat through the envelope in question is described by
the equation 
\begin{equation}
C {\partial \over \partial t}\ T - {\partial \over \partial z} 
\left( \kappa {\partial \over \partial z}\ T  \right) = Q,
\label{e:diff}
\end{equation}
where $T$ is the local (non-redshifted) 
temperature, $\kappa$ is the thermal conductivity, $C$ is the
heat capacity per unit volume at constant pressure, 
and $Q$ is the energy
generation rate per unit volume. Since the electron gas
is strongly degenerate, the heat capacities at constant volume 
and pressure are sufficiently close. We assume also that thermal processes do not
violate hydrostatic equilibrium ($\partial P/\partial z = \gs \rho$,
where $P$ is the pressure dominated 
by the relativistic degenerate electrons).

\begin{figure*}
\includegraphics[width=0.45\textwidth]{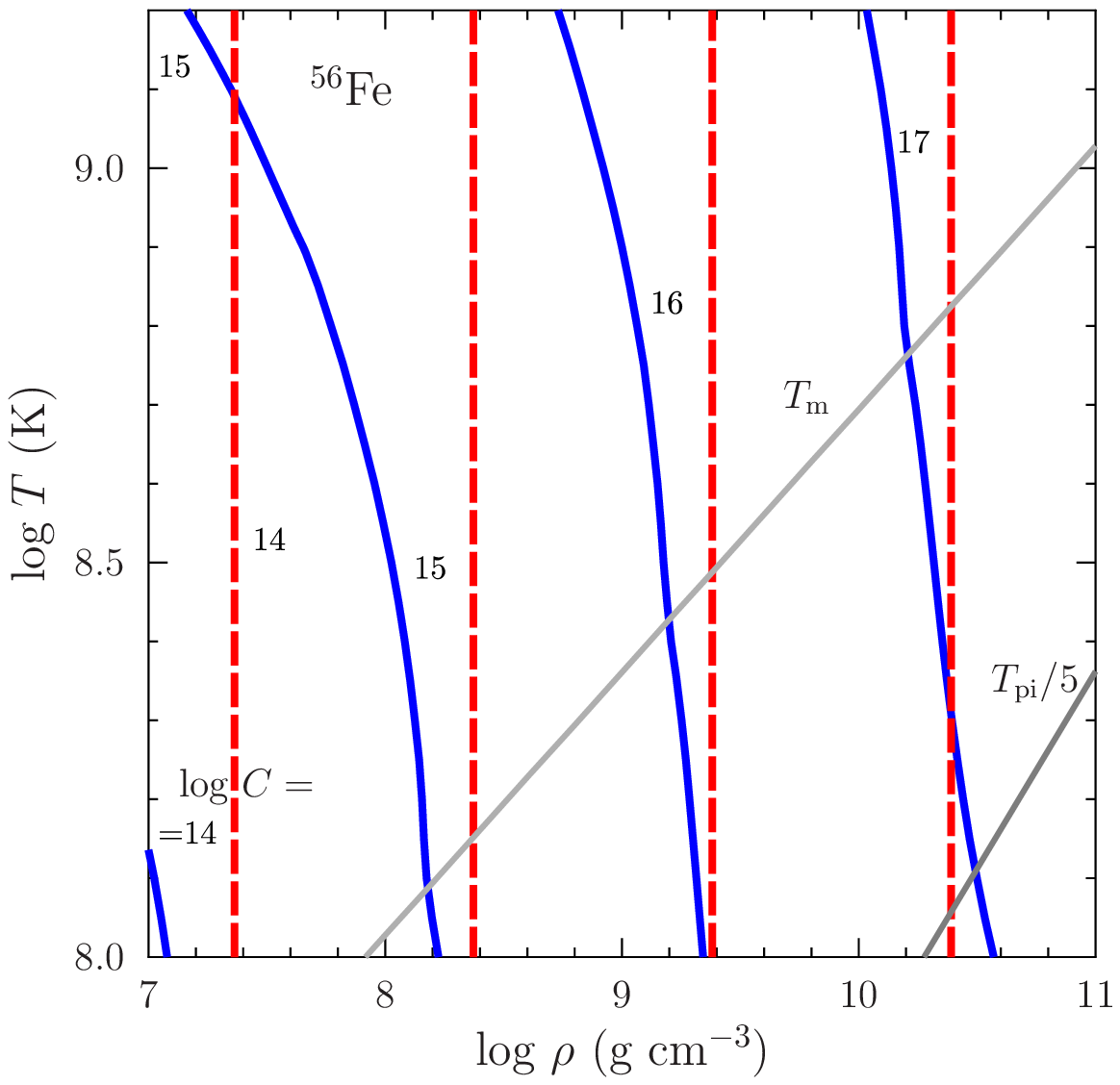}%
\hspace{5mm}
\includegraphics[width=0.45\textwidth]{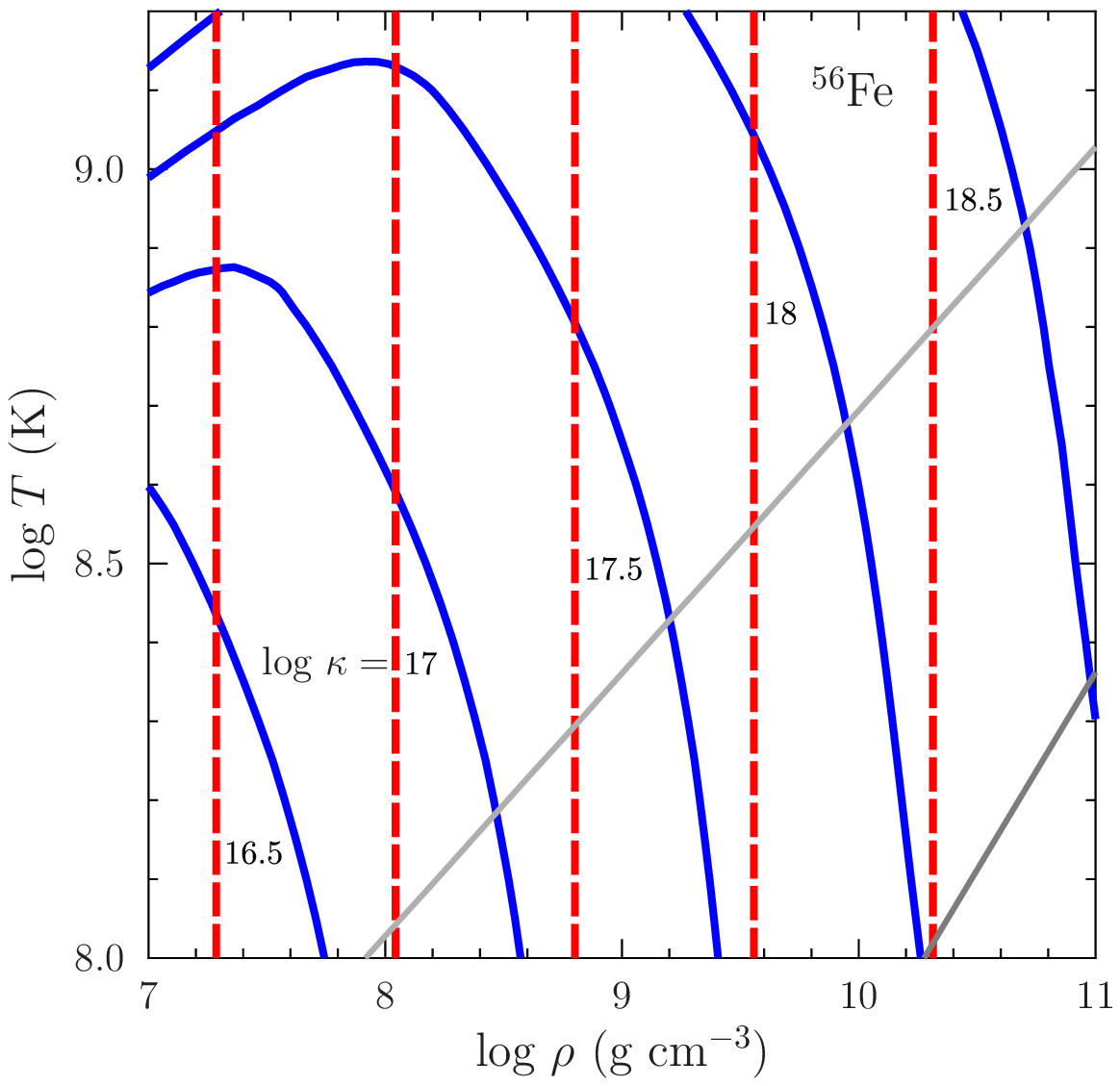}%
\caption{Isolines of constant heat capacity $C$ [$\kB$ cm$^{-3}$]
(the left-hand panel)
and thermal conductivity $\kappa$ [erg cm$^{-1}$ s$^{-1}$ K$^{-1}$] 
(the right-hand panel) in the $T-\rho$ plane
for iron matter. Solid lines are accurate values, dashed lines
refer to the toy-model approximation. 
The numbers show the values of log $C$ and log $\kappa$. 
See the text for details.
}
\label{f:Ckappa}
\end{figure*}

In reality, $C$, $\kappa$ and $Q$ depend both on density and on
temperature, which makes equation~(\ref{e:diff}) non-linear. In the
numerical simulations we take these dependences into account. In the toy
model, we linearize equation~(\ref{e:diff}) by assuming that $C$ and
$\kappa$ are temperature-independent (which is a reasonable
approximation, as we argue below) and that the dependence of energy
release on depth and time, $Q(z,t)$, is given explicitly.

In the given $\rho-T$ region (\ref{e:rho,T}) one can suggest two major
approximations of $C$ and $\kappa$. Since the electrons are strongly degenerate, the heat
capacity is mainly determined by the ions.  In an ideal classical crystal,
the ion heat capacity is $C_\text{i} = 3n_\text{i} \kB$, where $\kB$ is
the Boltzmann constant. Quantum effects strongly reduce  $C_\text{i}$ at
$T \ll T_\text{pi}$, where $T_\text{pi}=\hbar \omega_\text{pi}/\kB
\approx 7.8\times10^6\sqrt{\rho_6}\,(Z/A)$~K and $\omega_\text{pi}$  is
the ion plasma frequency. However, in real strongly coupled, strongly
degenerate non-ideal Coulomb plasma (liquid or crystal), the total heat
capacity per ion remains close to  $3\kB$ in a wide range of
temperatures around the melting line $T_\text{m} \sim 2.3\times10^7
(Z/26)^2 (A/56)^{-1/3}\rho_6^{1/3}$~K (e.g., \citealt{HPY2007},
Section~2.4.6). Bearing in mind an approximate nature of our analysis, we
take 
\begin{equation} 
C\approx 3 \kB n_\text{i}=a z^3, \quad
a \approx \frac{ \kB}{Z \pi^2}\, 
\left( \frac{\me c}{\hbar z_0} \right)^3,
\label{e:C}
\end{equation}
where we use $n_\text{i}=n_\text{e}/Z$ due to electric neutrality of the
matter.  In this approximation, $C$ is temperature-independent and
proportional to $\rho$.

The thermal conductivity $\kappa$ is mainly provided
by strongly degenerate electrons, which scatter off ions (off 
ion-charge fluctuations, to be exact). It is determined by the familiar
expression \citep[e.g.,][]{ZIMAN}
$\kappa=\pi^2 k_\text{B}^2 T n_\text{e} \tau_\text{eff}/(3 m_\text{e}^*)$,
where $\tau_\text{eff}$ is the effective electron relaxation time and
$m_\text{e}^*=m_\text{e}\,/\sqrt{1-v_\text{F}^2/c^2}$ is the effective
electron mass,
$v_\text{F}=c\xr/\sqrt{1+\xr^2}$ being
the electron Fermi velocity. For $\tau_\text{eff}$, we employ 
an estimate \citep{1980YU}, 
\begin{equation}
\frac{1}{\tau_\text{eff}}= \frac{e^2}{\hbar v_\text{F}}\,
\left(2 - \frac{v_\text{F}^2}{c^2}   \right)\, \frac{\kB T u_{-2}}{\hbar},
\label{e:taueff}
\end{equation}
where $u_{-2}\approx 13$ is a frequency moment of phonon spectrum
in a Coulomb crystal of ions. This estimate is obtained for electrons, 
which scatter off phonons at $T \gtrsim T_\text{pi}/5$. It neglects
quantum effects in ion motions and multi-phonon scattering processes
\citep{BKPY}. It stays roughly valid in a strongly
coupled Coulomb liquid of ions. In our case, it is sufficient 
to use the relativistic limit
($v_\text{F} \to c$), in which case
\begin{equation} 
\kappa \approx bz^2,\quad b=
\frac{\zeta \kB c^4 m_\text{e}^2}{9e^2 \hbar u_{-2} z_0^2}.
\label{e:kappa}
\end{equation}
Then $\kappa$ is independent of $T$ and $b$ is a constant. Here, we
introduce a phenomenological constant correction factor $\zeta$ which
makes our approximation of $\kappa$ more consistent with  
advanced calculations \citep{1999Palex}. For the iron plasma to
be considered below we set $\zeta=1/4$.

Substituting (\ref{e:C}) and (\ref{e:kappa}) into equation
(\ref{e:diff}), we have
\begin{equation}
a z^3 {\partial \over \partial t}\ T - b {\partial \over \partial z}
\left( z^2 {\partial \over \partial z}\ T  \right) = Q(z,t). 
\label{e:diff1}
\end{equation}
This is our basic toy-model equation, which is linear
in $T$ and can be solved by standard methods of mathematical physics
as we discuss later.

The accuracy of our approximations (\ref{e:C}) and (\ref{e:kappa}) 
is demonstrated in Fig.\ \ref{f:Ckappa}.  It shows  isolines 
of constant $C$ and $\kappa$ (the left-hand and right-hand
panels, respectively) in the $T-\rho$ plane. 
For illustration, here and below we 
use the model of the outer neutron star crust composed of iron. 
We have chosen iron as a leftover of nuclear burning
of light elements. 
The numbers next to the lines give the values of decimal logarithms
$\log C$ and $\log \kappa$. The vertical dashed lines are our approximations.
The solid lines are based on numerically accurate values of $C$ and
$\kappa$. Accurate $C$ includes the contribution of ions, electrons,
photons, as well as of electron-positron pairs.
Accurate $\kappa$ includes the contribution of electron-ion and
electron-electron collisions and also of radiative conduction. 
To guide the eye, the two gray lines show two characteristic temperatures (e.g.
\citealt{HPY2007})
as functions of $\rho$. The lighter line is the melting temperature $T_\text{m}$
of the classical Coulomb crystal of iron ions.
The darker line is $T_\text{pi}/5$. Below this line, 
quantum effects in ion motion substantially suppress
the heat capacity of the ions. 

According to Fig. \ref{f:Ckappa}, our approximations 
of $C$ and $\kappa$ seem reasonable.
Deviations of accurate and toy model heat capacities at
$\rho \lesssim 10^7$ \gcc\ and $T\gtrsim 3 \times 10^8$ K 
are mainly due to the contribution of electrons and photons 
in the accurate $C$ (the electron degeneracy becomes reduced 
which makes the electron and radiative heat capacities more important). 
The deviations at $\rho \gtrsim 3 \times 10^{10}$ \gcc\
and $T \lesssim 10^8$~K are due to quantum effects.
As for the accurate and approximate thermal conductivities,
their difference comes from the crudeness of our approximation (\ref{e:kappa}).
The accurate electron conductivity in the Coulomb liquid and crystal does depend on 
temperature \citep{1999Palex} although this dependence is not too strong in the
selected $T-\rho$ domain. At
$\rho \lesssim 10^7$ \gcc\ and $T\gtrsim 3 \times 10^8$ K the
radiative thermal conductivity becomes rather important.

To be specific, we take the star with $M=1.4$ \msun\ and $R=12$ km 
($g_\text{s}=1.59 \times 10^{14}$ cm~s$^{-2}$). Since the heat diffusion
in a thin outer stellar layer is self-similar \citep[e.g.][]{1983GPE}, one can easily
rescale to
other values of $M$ and $R$. In our case, we have $z_0=14.4$ m, 
$a=3.15 \times 10^3$ erg~cm$^{-6}$~K$^{-1}$ 
and $b=3.9 \times 10^8$ erg~cm$^{-3}$~s$^{-1}$~K$^{-1}$.
In order to rescale $a$ and $b$, it is sufficient
to notice that $z_0 \propto 1/g_\text{s}$ in equations (\ref{e:C})
and (\ref{e:kappa}).

\subsection{Analytic solution}
\label{s:analytic}

We apply equation (\ref{e:diff1}) for studying heat diffusion
from a heater (burst source), located in the outer crust, 
to the surface and to the stellar
interiors (to $z \to 0$ and $z \to \infty$, respectively).
We will use the solution at $z_\text{b} \leq z \leq z_\text{drip}$, where
$z_\text{b}$ and $z_\text{drip}$ correspond, respectively, to the densities
$\rhob$ and $\rho_\text{drip}$ in equation (\ref{e:rho,T}).

We present the solution as 
\begin{equation}
T(z,t)=T_0(z)+T_1(z,t),
\label{e:T0T1} 
\end{equation}
where $T_0(z)$
is a temperature profile in a quiet star (i.e., at $Q=0$), and
$T_1(z,t)$ is the temperature excess due to the burst;
$T_1(z,t)$ obeys the same equation (\ref{e:diff1}). 

The temperature profile 
$T_0(z)$ is determined by heat outflow from the neutron star
interiors ($z>z_\text{drip}$) and can be treated as stationary during
a burst and its successive decay. In our model,
equation (\ref{e:diff1}) with $Q=0$ yields
\begin{equation}
T_0(z)=T_\text{b0}+\frac{j_0}{b} \, \left( \frac{1}{z_\text{b}}-\frac{1}{z}
\right),
\label{e:T0}
\end{equation}
where $T_\text{b0}=T_0(z_\text{b})$, $j_0=\sigma_\text{SB}T_\text{s0}^4$ is the heat
flux emergent from stellar interiors; it is determined by the effective surface
temperature $T_\text{s0}$ (in the absence of the heater); $\sigma_\text{SB}$
is the Stefan-Boltzmann constant. The second term in equation (\ref{e:T0})
describes the steady-state temperature increase within the quiet star.

A solution of equation (\ref{e:diff}) for $T_1(z,t)$ is discussed
in Appendix. For an instant burst at  
$t=t_\text{h}$ in an infinitely thin 
shell located at $z=z_\text{h}$ 
we have $Q(z,t)=H_0 \,
\delta(t-t_\text{h})\, \delta(z-z_\text{h})$, $H_0$ being the energy
generated per unit area of the bursting shell. 
The solution for $t>t_\text{h}=0$ is
\begin{equation}
T_1(z,t)= \frac{H_0}{3bt \sqrt{z z_\text{h}}}\,
\exp \left(- \frac{u^2+u_\text{h}^2}{4t}   \right)\, 
I_{\frac{1}{3}}\left( \frac{u_\text{h}u }{2t} \right), 
\label{e:Green}
\end{equation}
where $I_{1 \over 3}(x)$ is a modified Bessel function
\citep[e.g.,][]{1953BE},
\begin{equation}
u=\frac{2}{3}\, \sqrt{\frac{a}{b}}\,z^{3/2},
\quad
u_\text{h}=\frac{2}{3}\, \sqrt{\frac{a}{b}}\,z_\text{h}^{3/2}.
\label{e:u}
\end{equation}

Equation (\ref{e:Green}) represents a Green's function to
equation (\ref{e:diff1}). It allows us to obtain a general
solution of equation (\ref{e:diff1}) with arbitrary
heat release distribution,
\begin{equation}
T_1(z,t)= \!\int\!\! {\rm d}z_\text{h} \,
{\rm d}t_\text{h} \frac{Q(z_\text{h},t_\text{h})}{3b t'
 \sqrt{z z_\text{h}}}
\exp \left(- \frac{u^2+u_\text{h}^2}{4 t'}   \right)
I_{\frac{1}{3}}\left( \frac{u_\text{h}u }{2 t'} \right), 
\label{e:Green1}
\end{equation}
where $t'=t-t_\text{h}$, and the integration is carried out
over the entire range of depths $z_\text{h}$ occupied by the heater and
over entire interval of times $t_\text{h}<t$, at which the heater 
is on at a given depth $z_\text{h}$. Equations {  (\ref{e:Green})} and (\ref{e:Green1}) 
allow fast computation 
of temperature evolution after any 
burst. Since our heat diffusion problem (\ref{e:diff1}) is linear, many
features of heat diffusion from the instant and thin heater apply for a
more general solution (\ref{e:Green1}).

\subsection{Toy bursts}
\label{s:toybursts}

The formulated model is restricted 
by the density and temperature range (\ref{e:rho,T})
and by neglecting neutrino cooling that becomes
significant at temperatures higher than a few
$\times10^9$ K
(e.g., \citealt{2004CM}).
For illustration, we consider toy bursts
with not very realistic parameters to stay in the
formulated parameter space.
We follow heat propagation after a burst
in the toy domain (\ref{e:rho,T})
using equation (\ref{e:Green1}).

\renewcommand{\arraystretch}{1.2}
\begin{table}
\caption{Two toy burst models A and B and their instant thin
counterparts ${\cal A}$ and ${\cal B}$ 
for a star with $M=1.4\,\msun$ and $R=12$ km; the ignition density is $\rho_2=10^8$ \gcc}
\label{tab:model}
\begin{tabular}{c  c c c c}
\hline 
Model$^{a)}$& $\rho_1$ $^{b)}$ & Model$^{c)}$  & $H_0^{d)}$  & $E_0$ $^{e)}$ \\
& \gcc\   &  & erg cm$^{-2}$ & erg  \\
\hline
A   & $ 3\times 10^7 $ & ${\cal A}$   &$5.02 \times 10^{26}$ & $9.08 \times 10^{39}$ \\
B   & $ 3\times 10^6$  & ${\cal B} $  &$5.99 \times 10^{26}$ & $1.08 \times 10^{40}$ \\
\hline
\end{tabular}
\\
$^{a)}$ Burning in a thick shell during $t_\text{burst}=100$ s at $\mathcal{Q}_\text{b}=5$ keV/N.\\
$^{b)}$ {  Lowest} burning density.
\\
$^{c)}$ Instant burning in an infinitely thin ignition shell ($\rho_1=\rho_2$).\\
$^{d)}$ Generated heat per 1 cm$^{2}$ column. \\
$^{e)}$ Total generated heat in the toy burst domain (\ref{e:rho,T}).
\end{table}
\renewcommand{\arraystretch}{1.0} 

We consider two toy finite-shell burst models denoted
as A and B (Table \ref{tab:model}). Their bottom (ignition)
density is fixed at $\rho_2=10^8$ \gcc. 
For burst A 
the top density 
of the burning shell
is $\rho_1=3 \times 10^7$ \gcc. The top density for
burst B, $\rho_1=3 \times 10^6$ \gcc, 
is taken lower than $\rhob$ to mimic standard models of  
superbursts as detailed in Section~\ref{s:thick}.

For bursts A and B, we assume the energy generation rate $Q(z,t)$ to
be proportional to the mass density $\rho(z)$ with
a fuel calorimetry $\mathcal{Q}_\text{b}=5$ keV per nucleon. This 
mimics burning of carbon mixed with a substrate (e.g. \citealt{2011KH}) in
our artificially weak superbursts. For simplicity, the fraction
of carbon in the heater before the burst is fixed,
so that $Q(z,t) \propto z^3$ and the main energy
release always occurs at the bottom  
of the heater (at $\rho=\rho_2$). As for the time dependence of
$Q(z,t)$, we assume that
the heater is switched on abruptly, operates at
a constant rate, and then it is turned off abruptly
as well. The duty time will be denoted as $t_\text{burst}$ and
set to be 100 s, for certainty. Note that the toy model
allows us to use any $Q(z,t)$ function, and we have tried
some versions in our test runs. 

Table \ref{tab:model}
lists also  the generated
heat per 1 cm$^2$ column, $H_0$, and the total energy 
$E_0=4 \pi R^2 H_0$ generated at the toy-model densities
$\rho\geq \rhob$.  

In addition, we will introduce simplified models (Table \ref{tab:model})
of instant bursts in infinitely
thin ignition shells ($\rho = \rho_2$), keeping the total burst energies the same.
We will mark them as ${\cal A}$ and ${\cal B}$.
Burst ${\cal B}$ is essentially the same as ${\cal A}$ 
but with slightly higher burst energy.

\subsection{Heat blanket and lightcurve}
\label{s:heat-blanket}

We use the toy model solution of free heat
diffusion after the burst in its applicability domain
(\ref{e:rho,T}). To calculate the effective surface temperature
$T_\text{s}(t)$ and the lightcurves for toy bursts we will treat the
outer layer  at $\rho < \rhob$ as the standard iron heat blanketing
envelope (e.g. \citealt*{1997PCY}), where the heat transport is
quasi-stationary and heat flux is conserved. Such envelopes are studied
separately; they establish a relation between $T_\text{b}$ and
$T_\text{s}$.

The heat blanket changes 
the heat diffusion regime at $\rho< \rhob$ and allows some heat to leak
to the surface and be observable 
as the surface emission. 
Such a scheme 
is justified if the heat blanket weakly affects the heat transport
under its bottom (see Appendix).
The inner boundary can be taken as isothermal at $z\to\infty$,
which is a good approximation for the considered burst parameters,
because $\rho_\text{drip}\gg\rho_2$.

We have calculated the dependence of $T_\text{b}$ on $T_\text{s}$ 
in the standard iron heat blanket, as in \citet{1997PCY},
and used it to obtain the
lightcurves by linking $T_\text{b}(t)$, calculated with the toy model,
to the
surface
luminosity $L(t)=4 \mathrm{\pi} R^2 \sigma_\text{SB}T_\text{s}^4$. 

Note that $z$ is a proper depth, $t$ is a proper time, and $L$ is a
non-redshifted luminosity (for a local observer). The redshifted 
(Schwarzschild) time $t_\text{S}$ and luminosity $L^\infty$  (for a
distant observer) are given by \citep*[e.g.,][]{MisnerTW}
\begin{equation}
t_\text{S}=\frac{t}{\sqrt{1-r_\text{g}/R}}, \quad
L^\infty= (1-r_\text{g}/R)\,L.
\label{e:infty} 
\end{equation}

According to equation (\ref{e:T0T1}), 
$T_\text{b}(t)=T_0(z)+T_1(z_\text{b},t)$,
where $T_0(z)$ is the temperature prior to the burst. 
For simplicity, we will often assume that $T_0$ is much
smaller than the characteristic excess temperature $T_1$ during the burst
($T\approx T_1 \gg T_0$), and 
the luminosity $L_0$ prior to the burst is
much smaller than $L$. We will call this the $T_0 \to 0$ approximation. 
In some cases, to be more realistic, we will set
$T_\text{b0}=10^8$ K. Then the surface temperature
and thermal luminosity prior to the burst are $T_\text{s0}=9.73 \times 10^5$ K and
$L_0=9.2 \times 10^{32}$ erg~s$^{-1}$, respectively.

\subsection{Numerical simulations}
\label{s:check}
    
Since we do not expect the toy model to be very accurate, we will check
its results with a few test runs done with a numerical code of
neutron-star thermal evolution. Such simulations would be inappropriate
while using the standard cooling codes (e.g.,
\citealt{2014KAM,2018Chaikin}), which assume the stationary temperature
profiles at $\rho < \rho_\text{b}$   and
barotropic equation of state (that is, $T$-independent pressure) due to
the strong degeneracy at $\rho>\rho_\text{b}$. The shallower layers at
$\rho<10^7$ g cm$^{-3}$ can be essentially non-stationary at the
timescales of hours and days that we consider in the present work. The
relaxation time of the envelope could be made shorter by shifting
$\rho_\text{b}$ to lower densities, but such densities cannot 
be accurately modeled
by the standard cooling codes because the matter is not strongly
degenerate and the equation of state is not barotropic. We perform the
simulations  using the numerical code described in \citet{PC18}, which
is free from the above assumptions. It allows us to get rid of a
relatively thick quasi-stationary heat-blanketing envelope, required in
the standard cooling codes, and to treat evolution of non-degenerate and
partially degenerate layers of the star on equal footing with the
strongly degenerate interiors.  The code employs modern microphysics
(see \citealt*{2015PPP} for a review). The hydrostatic equilibrium and
heat transport equations are solved consistently, using the number of
baryons inside a given shell as an independent variable
\citep*[cf.][]{Richardson_79}. This code still uses an outer
quasi-stationary envelope to simplify the treatment of the zone of
partial ionization, but the choice of the boundary is more flexible. In
this case, the density at the bottom of the outer envelope $\rhob$ would
be an inadequate parameter, because it depends on $T$. The
temperature-independent parameter that we actually use is the baryon
mass of the outer envelope $M_\text{env}$. 

The heating and cooling simulations were performed for  $M=1.38\,\msun$
neutron star using the BSk26 model of the equation of state and
composition of the inner crust and the core \citep{Pearson_18}. Having
the radius $R=11.83$ km, this star has the same compactness
$r_\text{g}/R$ and almost the same surface gravity $g_\text{s}$ as our
basic 1.4 \msun\ star with $R=12$ km. The outer crust is assumed to
contain only iron ions, as in the toy model. For the outer envelope we
have chosen $M_\text{env}=10^{-12}\,\msun$, $10^{-13}\,\msun$ or
$10^{-14}\,\msun$. At low temperatures, these choices roughly correspond
to $\rhob\sim10^6$ \gcc, $2\times 10^5$ \gcc{} or $5\times10^4$ \gcc,
respectively.

The microphysics of deep stellar layers (at $\rho\gg\rho_2$) has no
direct effect on bursts A and B. Before the burst, the quasi-equilibrium
temperature profile with $T=10^8$~K at  $\rho=10^7$ \gcc{} was selected
from the neutron-star cooling sequence. In this case the core is almost
completely isothermal. Because of its large heat capacity and high
thermal conductivity, the core keeps a constant temperature on the
timescales under consideration (during the burst and afterburst
relaxation of the crust). Therefore the details of the core microphysics
(composition, superfluidity, neutrino emission mechanisms etc.) are
unimportant in the present study. In the simulations, we take the same
heating power $Q(z,t)$ as in the toy bursts A or B. Below we will
compare the computed temperature profiles and lightcurves with  the toy
models.      

\subsection{Short nuclear burning phase}
\label{s:instant-heater}

Outbursts in neutron star crust are complex phenomena with a number of 
different time scales. The shortest time scale in our consideration is
the nuclear energy release (taken to be $t_\text{burst}=100$ s).  
It is so short that the fraction of heat that escapes from
the burst area during this time is negligible. For this reason,
its exact duration is insignificant for further thermal evolution of
bursts A or B; it is  the total  generated heat that really matters.
Both the numerical and toy models describe the
temperature evolution during the energy release. We will follow this
evolution but will not focus on this phase.

It is important to note that in bursting neutron stars one often uses
(e.g.\ \citealt{2004CM,2006Cumming,2012Altamirano}) the approximation of
instant heater to describe the initial temperature rise in the burning
layer.  This approximation assumes  instant transformation of the
nuclear energy into heat in  any element of the burning layer,
neglecting  heat transport mechanisms. Then the temperature jumps from
its initial values  $T_0(\rho)$ to the values $T_\text{f}(\rho)$, which are
determined solely by the sudden local heating. {  These values} are controlled  by the
heat capacity and nuclear energy release. This approximation  allows
one  to skip the initial fast temperature rise, {  which saves} computer time.

Since the toy-model  assumes the classical ion heat capacity, equal to
$3\kB$ per a nucleus, instant burning gives the excess temperature  jump
$T_{1\text{f}}(\rho)=A\mathcal{Q}_\text{b}/(3 \kB)$ in the  burning layer, with
$T_{1\text{f}}=0$ outside this layer.  With $A=56$ and $\mathcal{Q}_\text{b}=5$ keV per nucleon,
we have one and the same constant temperature  jump $T_{1\text{f}}(\rho)=1.08
\times 10^9$ K within the heater for burst models A and B in Table
\ref{tab:model}.  The constancy of the toy-model $T_{1\text{f}}(\rho)$  results
from constant heat capacity per baryon.  {  Our} numerical simulations use
more realistic microphysics with higher heat capacity at sufficiently
low $\rho$ and high $T$ (Fig.~\ref{f:Ckappa}), mainly due to a
contribution of the electrons, which are less degenerate at lower
densities. Accordingly, the simulations predict lower $T_\text{f}(\rho)$ 
{  (compared with the toy model)} and rising $T_\text{f}(\rho)$ profiles in the
burning zones, as will be 
discussed in Sections \ref{s:burstAA} and \ref{s:thick}
below (cf., e.g., \citealt{2004CM}). One can also change
the $T_\text{f}(\rho)$  profile assuming density dependent  fraction of nuclear
fuel within the burning layer (for instance,  due to nuclear evolution
prior to burst or incomplete burning  during the burst; e.g.
\citealt{2004CM,2006Cumming,2015Keek}).

The initial temperature rise in the idealized promptly  bursting shells
${\cal A}$ and ${\cal B}$ (Table \ref{tab:model}) is different. For an
instant burst in an infinitely thin spherical shell, the initial
temperature rise $T_\text{f}$ is a delta-function, which is infinite at the
burst moment and at the shell location ($t=0$, $\rho=\rho_2$). It is
smoothed out later by heat transport.  

\subsection{Three burst stages (I, II, III)}
\label{s:stages}

After the short burning phase, one often distinguishes three 
burst stages which we denote as stages I, II, and III. These stages
have been described in the literature 
(e.g.\ \citealt{2004CM,2006Cumming,2012Altamirano}).

Stage I is characterized by a strong initial dynamical heat transport above the 
ignition layer (at $\rho<\rho_2$),
corresponding to an increase of the output heat flux over time. 
It ends after the onset  
of a slowly time-varying  { ({\it quasi-stationary})}
heat outflow in the outer layers. This stage is followed by stage II of
most energetic energy release through the surface. 
During
stage II  
the regime 
of quasi-stationary heat outflow 
establishes everywhere above
the ignition layer. 
The final stage III of burst decay is realized
when the generated heat starts to sink predominantly inside the star;
it corresponds to a reversal of the heat flux
in the toy model.
We will describe these stages for our models below.

\section{Instant thin-shell bursts}
\label{s:thin-instant}

We start with the simplest idealized instant ($t_\text{burst}=0$) 
toy-burst ${\cal A}$ in the infinitely thin shell at $\rho_2=10^8$ \gcc,
but with the same total energy release as in the more realistic model A
(Table~\ref{tab:model}).

\subsection{Temperature profiles and lightcurve}
\label{s:TrhoAA0}

\begin{figure}
\includegraphics[width=0.45\textwidth]{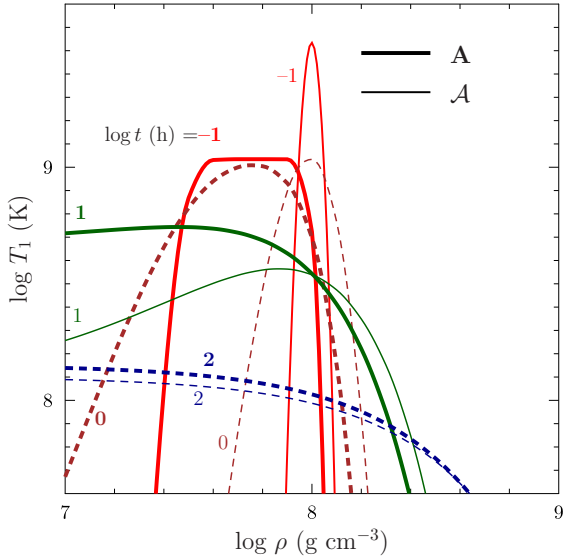}%
\caption{Excess temperature  profiles versus density at four 
moments of time $t$=0.1, 1, 10 and 100 h (marked by $\log t$ [h])
after an instant toy burst ${\cal A}$ 
in the thin shell at $\rho=10^8$ \gcc\ with the total
energy release $E_0=9.08 \times 10^{39}$ erg (thin lines). 
Thick lines are the same but produced by toy burst A from the shell 
of finite thickness. See the text for details. 
}
\label{f:TrhoAA0}
\end{figure}

The excess temperature profiles $T_1(\rho)$ are given 
by equation (\ref{e:Green}).
Fig.\ \ref{f:TrhoAA0} shows  
these profiles (thin lines) produced 
in the outer crust of the neutron star
after burst ${\cal A}$ in an ignition shell  
with the total energy release $H_0=9.08 \times 10^{39}$ erg;
$T_1$ is plotted 
as a function of density 
at four moments of time $t$ since the burst starts, 
$t$= 0.1, 1, 10 and 100 hours. Thick lines show similar profiles for burst A
in the shell of finite thickness.

An initial delta-function temperature spike becomes lower, wider and asymmetric 
and then disappears as the heat spreads over the crust. 
In this model, stage I lasts for about 10 h during which the thermal wave moves 
predominantly to the surface and
reaches the outer layers ($\rho \sim 10^7$ \gcc). Stage II lasts till
$t \sim (30-40)$~h. By this time the $T_1(\rho)$ profile becomes 
nearly horizontal above the ignited shell. 
Heat diffusion  
{  slows down, which suppresses} the heat flow to the surface. At the last
stage III the internal thermal wave moves slowly inside the star beyond the
ignition shell.

\begin{figure}
\includegraphics[width=0.45\textwidth]{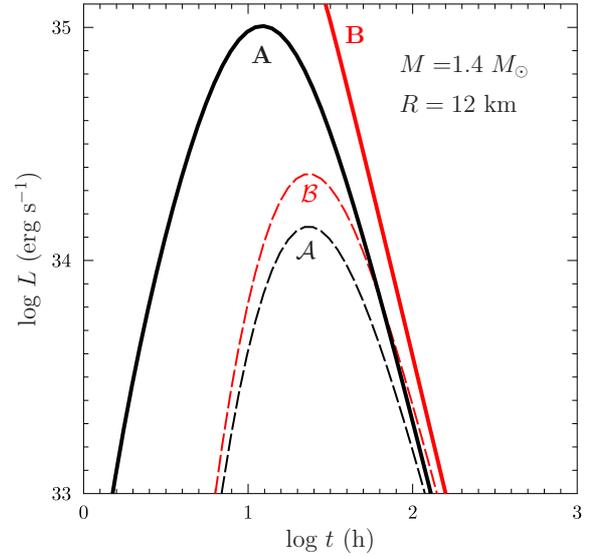}%
\caption{Thick lines show burst luminosity versus time 
for toy-burst models A and B (Table \ref{tab:model}) assuming $T_0 \to 0$. 
The segments of especially high $L(t)$ are not shown because 
the  toy-model cannot reproduce them accurately. 
Dashed lines are the lightcurves for the associated models 
${\cal A}$ and ${\cal B}$  of instant bursts in the ignition
shell ($\rho=\rho_2$). See the text for details.
}
\label{f:l(t)}
\end{figure}

The appropriate lightcurve for burst ${\cal A}$ is plotted in Fig.\ \ref{f:l(t)} along
with the lightcurves for other burst models. 
The lightcurve reaches its peak in $t \approx 20$ hours when the most
energetic part of the thermal wave emerges at the surface. Later the
lightcurve decays; {  the decay} is nearly power-law at the final stage.

Actually, the temperature profiles $T_1(\rho)$ at the early stages of
the instant ignition-shell burst models contain 
rapidly increasing segments which
are unstable against convection. The convection can change the temperature
profiles and the early segments of lightcurves, which we discuss below
for more realistic burst model~A.

\subsection{Basic properties of instant bursts}
\label{s:propertiesAA0}

Equation (\ref{e:Green}) possesses the following properties.

Firstly, in a small vicinity near the heater ($|z-z_\text{h}|\ll z_\text{h}$) just
after the heat release 
equation (\ref{e:Green})
reduces to
\begin{equation}
T_1(z,t)= \frac{H_0}{2 C_\text{h} \sqrt{\pi \mathcal{D}_\text{h} t}}
 \exp \left( - \frac{(z-z_\text{h})^2}{4\mathcal{D}_\text{h}
t} \right).
\label{e:GreenA}
\end{equation}
where $\mathcal{D}_\text{h}=b/(a z_\text{h})$
 and $C_\text{h}=a z_\text{h}^3$ have meaning of
the diffusion coefficient and the heat capacity near the heater,
respectively. This is the well known temperature
distribution produced after a  point-like and instant heat release  in a
uniform medium. Accordingly, just after the burst one half of the 
thermal energy diffuses to $z \to 0$ while the other half diffuses to $z
\to \infty$.

Secondly, it is easy to show that if a neutron star 
crust consisted solely  of the toy-model 
matter down to the surface $z=0$, 
all the heat generated 
within the crust 
would diffuse, 
on a long run, 
within
the star. No heat would be
able to flow through the surface because the
toy thermal conductivity (\ref{e:kappa})
vanishes at $z=0$. The initial heat outflow to the surface would be 
redirected later inside the star (Fig.\ \ref{f:TrhoAA0}). 
This is the basic 
heat-{accumulating} property 
of inner layers of the outer neutron star crust. 
This possibility of heat {accumulation} in the crust has been
pointed out by \citet{ec89}. Since the heat diffusion
problem is linear in the toy model, the heat propagation from extended
heaters would possess the same property.

The heat diffusion described formally by equation (\ref{e:Green}) to $z=0$
would give finite  $T_1(t)$ but zero heat flux at $z=0$. 
This excess temperature $T_1(t)$ would grow up when the thermal wave 
reaches the surface; it would fall down later when the 
heat would start sinking 
inside
the star after reflecting off the
absolutely insulating surface.   

This  unphysical behaviour is caused by the formal extension of the toy
model to $z\to0$, discussed in the Appendix. 
Actually,
the physical assumptions
underlying the toy model are only justified in the $\rho-T$ domain
(\ref{e:rho,T}). Therefore, in the figures we only show the results
obtained using the toy model at $\rho>10^7$ \gcc.
Microphysics in the outer layer ($\rho \lesssim 10^7$ \gcc) is different
and allows some heat to outflow through the surface, which we approximately describe
by  introducing the heat blanketing envelope 
(Section \ref{s:heat-blanket}).

Even in the selected domain (\ref{e:rho,T}) the toy model may somewhat
exaggerate the announced heat {  accumulation} (due to the neglect of
quantum suppression of heat capacity of crystalline ions) 
or to underestimate it (because of the neglect of the
electron contribution to the heat capacity). Nonetheless, we believe
(and confirm by the numerical simulations) that the model adequately
reflects this heat {  accumulation} and enables one to study its consequences.   

Note that at $t \gg (u^2+u_\text{h}^2)/4$
from equation (\ref{e:Green}) we
have
\begin{eqnarray}
T_1(z,t)&=&\frac{H_0}{3b\sqrt{z_\text{h} z}\, t\, \Gamma(4/3)}\, \left(
\frac{u_\text{h}u }{4t} \right)^{1/3}
\nonumber \\
&=&
\frac{H_0}{3b\Gamma(4/3)} \left(\frac{a}{9b} \right)^{1/3}
\frac{1}{t^{4/3}},
\label{e:Greenasy}
\end{eqnarray}
where $\Gamma(4/3)\approx0.893$ is the gamma-function value. 
In this case, $T_1(z,t)$ becomes independent of $z$ and $z_\text{h}$ and decreases with
$t$ as $t^{-4/3}$, determining the very late asymptotic behaviour 
of the lightcurve $L(t)$.

\section{Finite-width shell burst A}
\label{s:burstAA}

\subsection{Overview}
\label{s:overview}

Here we discuss burst A (Table \ref{tab:model})
in a sufficiently thick spherical layer [$\rho=(3 -10)\times 10^7$ \gcc] 
that fully lies
within the toy-model density range (\ref{e:rho,T}). Model ${\cal A}$, that
has been analysed in Section \ref{s:thin-instant}, 
represents a thin-shell counterpart of model~A. 

The thick lines in Fig. \ref{f:TrhoAA0} show snapshots of the toy-model 
A excess temperature
profiles $T_1$ versus $\rho$ at different moments of time
in comparison with burst ${\cal A}$ 
(thin lines). The $T_1(\rho)$ curves can be regarded as the $T(\rho)$ curves
in the $T_0\to 0$ approximation.
We see that the A and 
${\cal A}$ profiles of $T_1(\rho)$ in Fig. \ref{f:TrhoAA0}
are different at $t \lesssim 1$ d. The largest difference
is just after the burst 
(with flat $T_\text{f}(\rho)$-profile within the heater 
for burst A versus sharp spike for burst ${\cal A}$)
but they become
close later. The appropriate lightcurves can be compared in Fig.\
\ref{f:l(t)} with the same conclusion.

We have checked that the internal
temperature profiles and lightcurves at the late stage III (Section \ref{s:stages}) of 
burst model B (Table~\ref{tab:model}) are well described by the respective
model ${\cal B}$. This seems to be a generic feature
of bursts associated with the fact that the main burst energy is released
near the ignition density $\rho_2$.

\begin{figure*}
\includegraphics[width=0.45\textwidth]{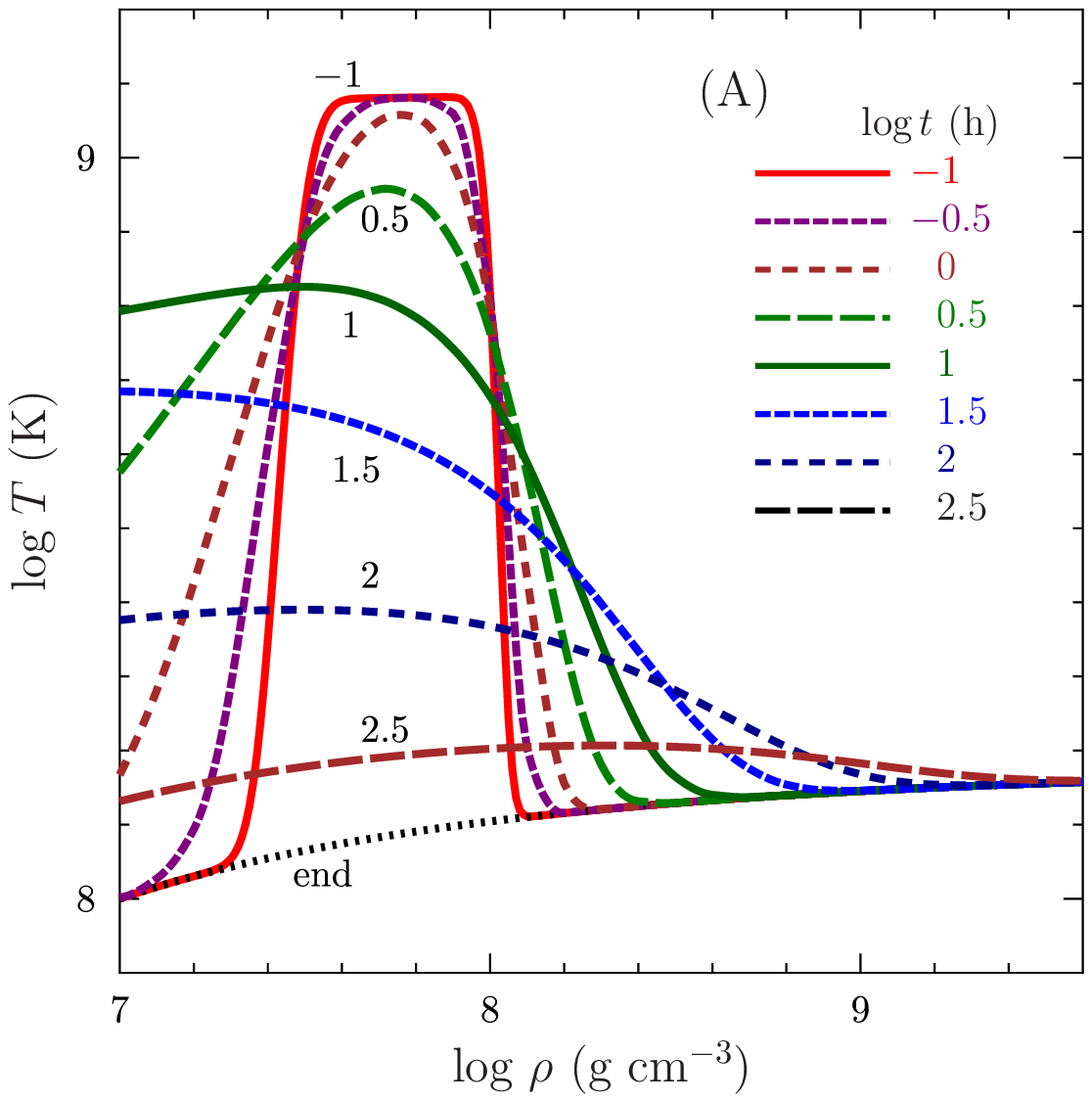}%
\hspace{1em}
\includegraphics[width=0.46\textwidth]{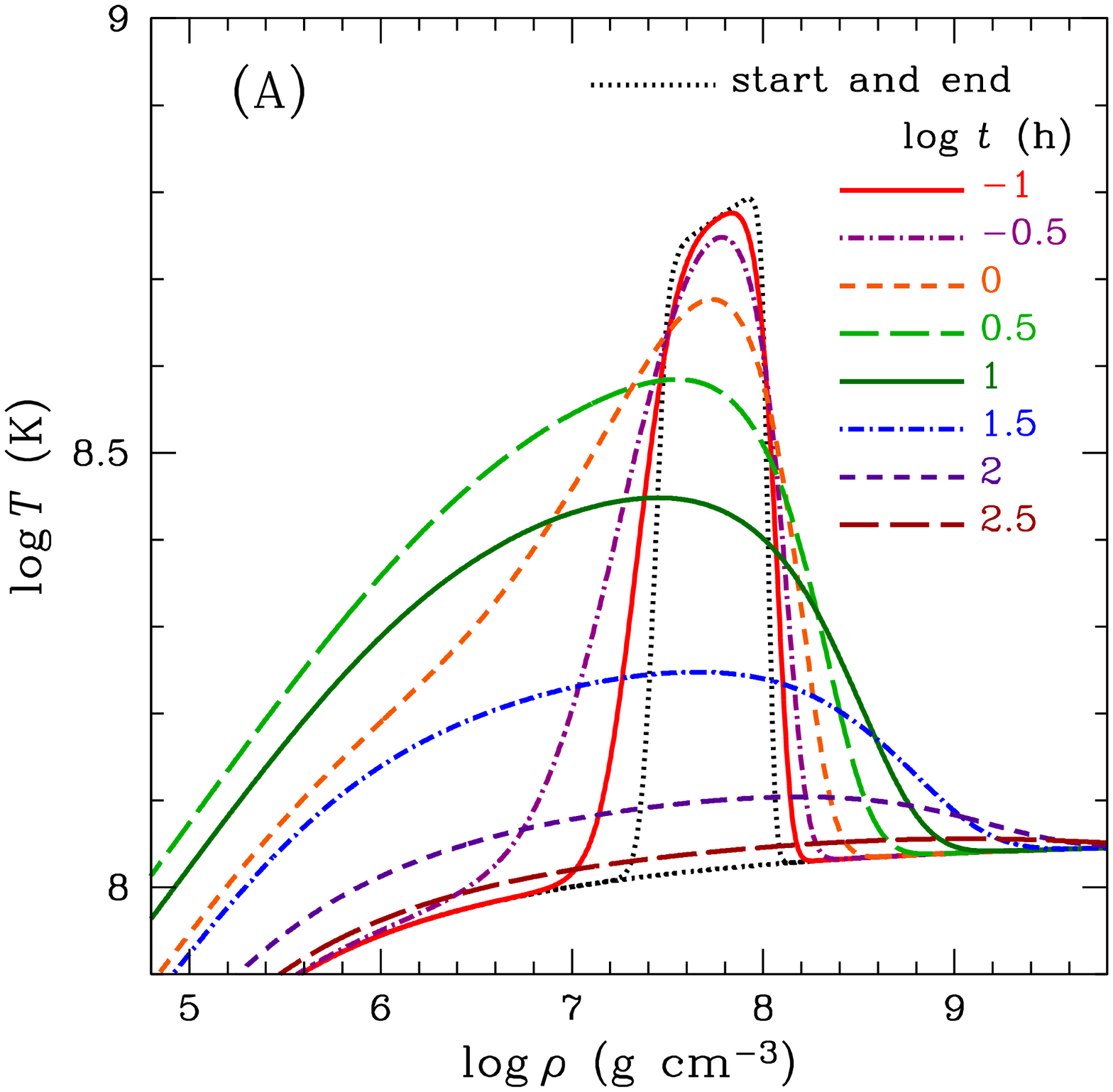}%
\caption{Density dependence of the internal temperature $T(\rho)$ 
in the outer neutron star crust   
after burst A 
at different moments of time $t$ (marked by $\log t$ [h]) assuming 
$T_\text{b0}=10^8$ K at $\rho=10^7$ \gcc\ prior to the burst.
The curves on the left-hand panel are calculated with the toy model,
while those on the right-hand panel are computed with the numerical
code that allows us to extend them to lower $\rho$.
The lower dotted curves on both panels are pre-burst (= after-burst) 
temperatures $T_0(\rho)$.
See the text for details.
}
\label{f:TrhoAA}
\end{figure*}

Fig.~\ref{f:TrhoAA} presents the internal temperature profiles
$T(\rho)$ at different moments of time $t$ (marked by the values of
$\log t$ [h]) after burst A assuming the pre-burst  temperature
$T_\text{b0}=10^8$ K at $\rho=10^7$ \gcc. The curves on the left-hand
panel are calculated using the toy model while the curves on the
right-hand panel are calculated by the thermal evolution code. The
lower dotted line is the temperature profile $T_0(\rho)$ without any
burst [it is given by equation (\ref{e:T0}) for the toy model]. As
explained in Section~\ref{s:check}, the code allows us to compute the
$T(\rho)$ profiles at any densities. Here the
outer envelope with mass $M_\text{env}=10^{-14}\,\msun$ is used, which enables us
to display the $T(\rho)$ curves to lower densities $\rho\sim10^5$ \gcc{}
in the right-hand panel.

\begin{figure*}
\includegraphics[width=0.45\textwidth]{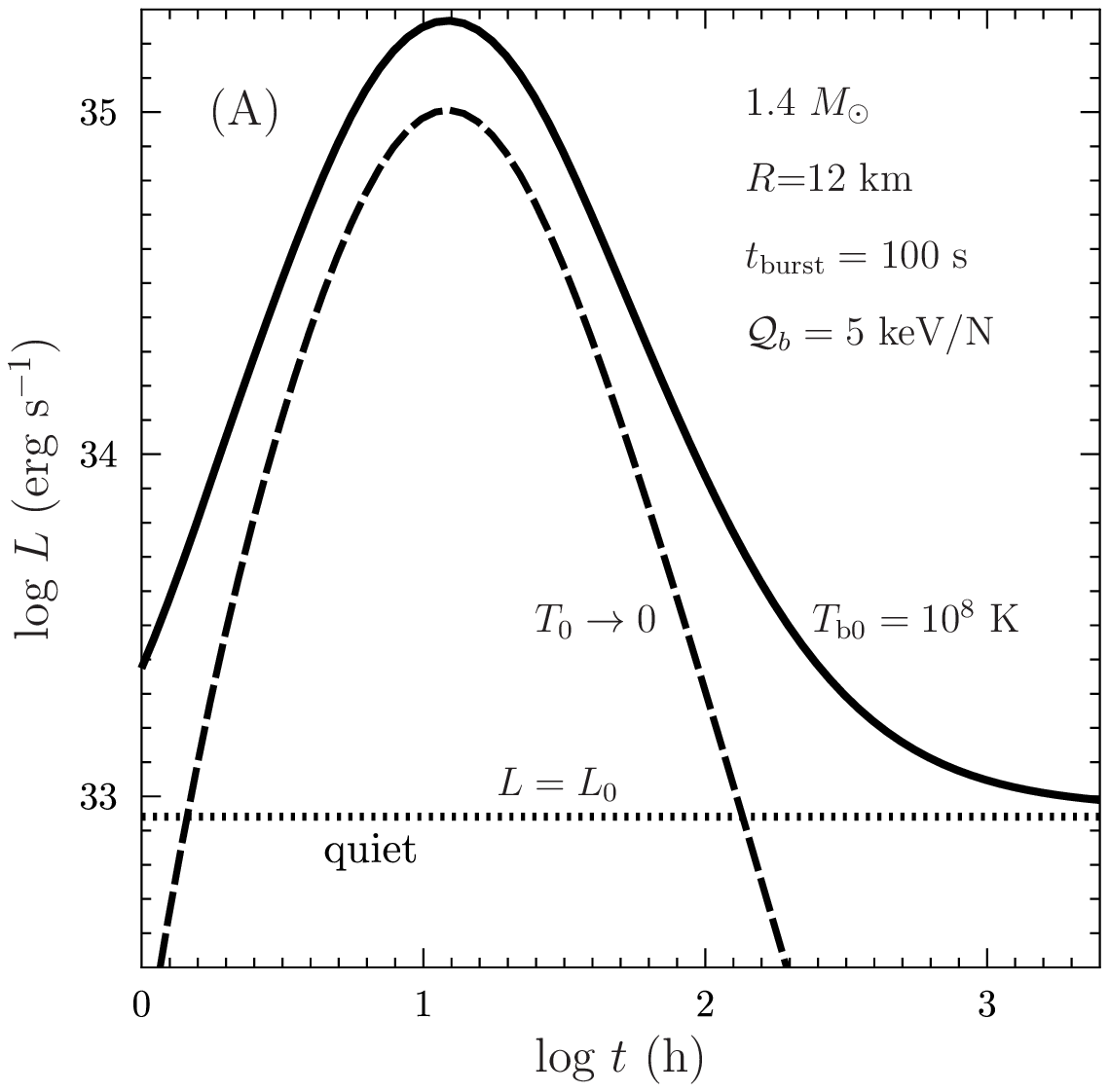}%
\hspace{5mm}
\includegraphics[width=0.45\textwidth]{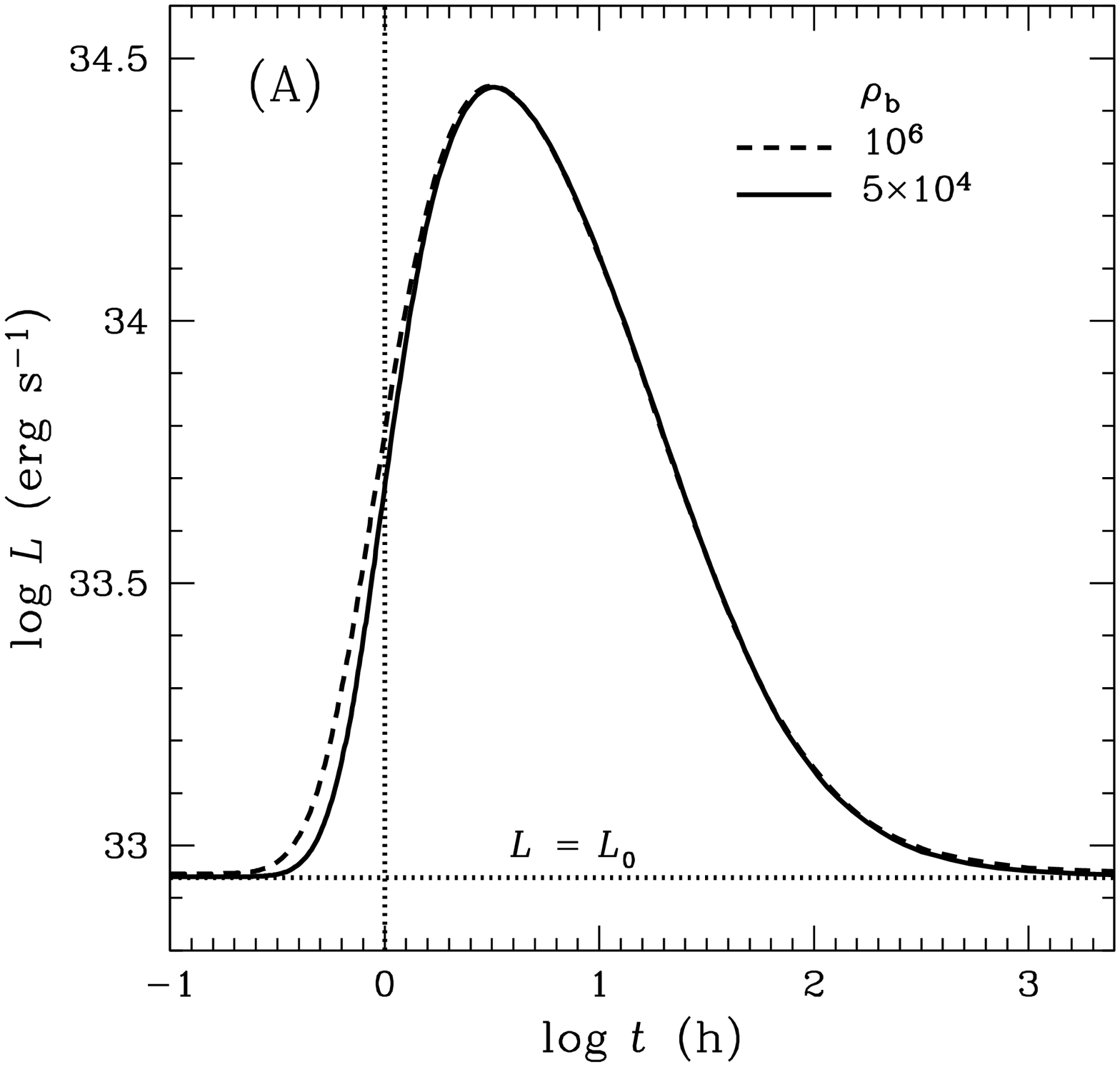}%
\caption{Surface luminosity versus time for burst A  calculated  with
the toy model (left) and with the thermal evolution code (right).  The
solid line on the left-hand panel as well as solid and dashed lines on
the right-hand panel correspond to $T_\text{b0}=10^8$ K at $\rho=10^7$
\gcc\ prior to the burst; the horizontal dotted lines display the
quiescent  luminosity. The dashed line on the left-hand  panel shows the
toy lightcurve assuming $T_0 \to 0$ (as in Fig.\ \ref{f:l(t)}). The
solid and dashed lines on the right-hand panel are computed with
different heat blanket models. The vertical dotted line in the right-hand
panel is a guide to the eye for comparison with the solid line in the left-hand panel.
See text for details.
}
\label{f:LAA}
\end{figure*}

Fig.~\ref{f:LAA} presents the lightcurves for burst A calculated using
the toy model (on the left-hand panel) and the thermal evolution code
(on the right-hand panel). The solid curve in the left-hand panel
and both curves in the right-hand one
refer to $T_\text{b0}=10^8$ K at $\rho=10^7$
\gcc; the corresponding levels of the quiescent thermal luminosity of
the star are plotted by the horizontal dotted lines. The dashed line for
the toy model presents the lightcurve assuming $T_0 \to 0$ (as in Fig.\
\ref{f:l(t)}). The solid and dashed lines on the right-hand panel are
computed for the same burst model but using different heat blankets
(with equivalent $\rhob=5 \times 10^4$ and $10^5$ \gcc, respectively).
The nice  agreement between these curves shows that the outer
quasi-stationary envelope of $M_\text{env}=10^{-14}\,\msun$ is
sufficiently thin to ensure good accuracy of the simulations.

According to Fig.\ \ref{f:l(t)}, burst A becomes pronounced in
the surface emission in a few hours after the explosion, in contrast with
$\sim 10$ hours for burst ${\cal A}$. This is because the outer
part of the burning layer A  is closer to the surface. 

\subsection{Burst A: toy model versus simulations}
\label{s:toyvscodeAA}

Now we can compare the toy-model results with those provided by the
numerical simulations. Figs.~\ref{f:TrhoAA} and \ref{f:LAA} allow
us to compare $T(\rho)$ profiles and the lightcurves of burst A. 

The overall qualitative agreement seems reasonable although some differences
are substantial. The  
differences are visible at stage I which
lasts for a few hours 
and
at stage II, that ends
in about 30 hours. The agreement between the toy and accurate results 
at the last decay stage III is more satisfactory.

The main source of disagreement is in the underestimation of the heat
capacity at $\rho \sim 10^7$ \gcc\ in the toy model (as discussed above;
Fig.\ \ref{f:Ckappa}) and much more realistic treatment
of the heat transport to the very surface by the numerical code. With
the reduced toy heat capacity,
 the instant-afterburst toy temperature 
$T_\text{f}(\rho)$ (Section~\ref{s:instant-heater}) in the burning zone 
becomes higher than it should be. These
instant afterburst segments of the $T_\text{f}(\rho)$ curves are quite visible in
Fig.~\ref{f:TrhoAA} (at $\log t\ [{\rm h}] = -1$ and $-0.5$).
The largest $T_\text{f}$ difference reaches a factor 
$\sim (2-3)$ 
at $\rho=3 \times
10^7$ \gcc. As a result, the toy model overheats the matter
at lower densities, making the lightcurve noticeably brighter than
it should be at stages I and II. It overestimates the burst energy radiated
at these stages through the surface and reduces in this way heat-accumulating  
properties of neutron stars. Owing to these reasons we do not show
most luminous segments of the toy lightcurve  B in 
Fig.\ \ref{f:l(t)}. According to the simulations, about 20 per cent
of the burst energy emerges through the surface in burst A. The toy model
does not allow us to accurately estimate this value.    

Let us 
note sufficiently large temperature gradients
of the toy $T(\rho)$ profiles (Fig.\ \ref{f:TrhoAA}) near
$\rhob=10^7$ \gcc\ at stage I. They are expected to be badly compatible with
the toy heat blanket model (Section \ref{s:heat-blanket}) making the
toy lightcurve even less reliable. Note also that the quiescent (dotted)
$T(\rho)$ profile is steeper for the  toy model.

On stages
II and III, both approaches predict nearly horizontal segments of 
the $T(\rho)$ profiles  
(with small inclinations relative to the horizontal axis)
which correspond to quasi-stationary 
and nearly
flux-conserving heat propagation. 
These segments appear rather insensitive 
to microphysics of the matter:
the heat capacity drops out of equation (\ref{e:diff}) in the stationary
case and the thermal conductivity should only be high enough to
ensure almost horizontal profiles.

\subsection{Convection after burst A}
\label{s:convection}

Steeply rising segments of the $T(\rho)$ profiles for 
bursts A and ${\cal A}$ at stage I in Figs.~\ref{f:TrhoAA0} and
\ref{f:TrhoAA} can be convectively
unstable.
The convection has been neglected
both in the toy model and in the numerical simulations. Let
us outline it for the toy model after burst A. To estimate the deepest densities of
the convective zone, we have used 
accurate microphysics of fully ionized plasma of iron matter. 
We have employed the Schwarzschild
convection criterion and compared the toy-model temperature gradients with the
adiabatic ones.

As a result,
we have obtained that the convection
can operate at stage~I for about 5 hours after the burst. Later
the bottom density of the convective zone
becomes lower than $\rhob=10^7$ \gcc, and the convection 
disappears from the toy-model domain (\ref{e:rho,T}). It can still operate at
$\rho<\rhob$,
but it cannot strongly affect {  the model lightcurves
and heat propagation at $\rho>\rhob$}. 

If the convection is on, the real $T(\rho)$ profiles lie between the
heat-diffusion and adiabatic temperature profiles, and the latter can be
essentially higher
than the former. We do not plot the adiabatic $T(\rho)$ curves 
and we do not follow the consequences
of convection in detail because we regard models A and ${\cal A}$ as illustrative.

\section{Thick-shell burst B}
\label{s:thick}

\begin{figure*}
\includegraphics[width=0.45\textwidth]{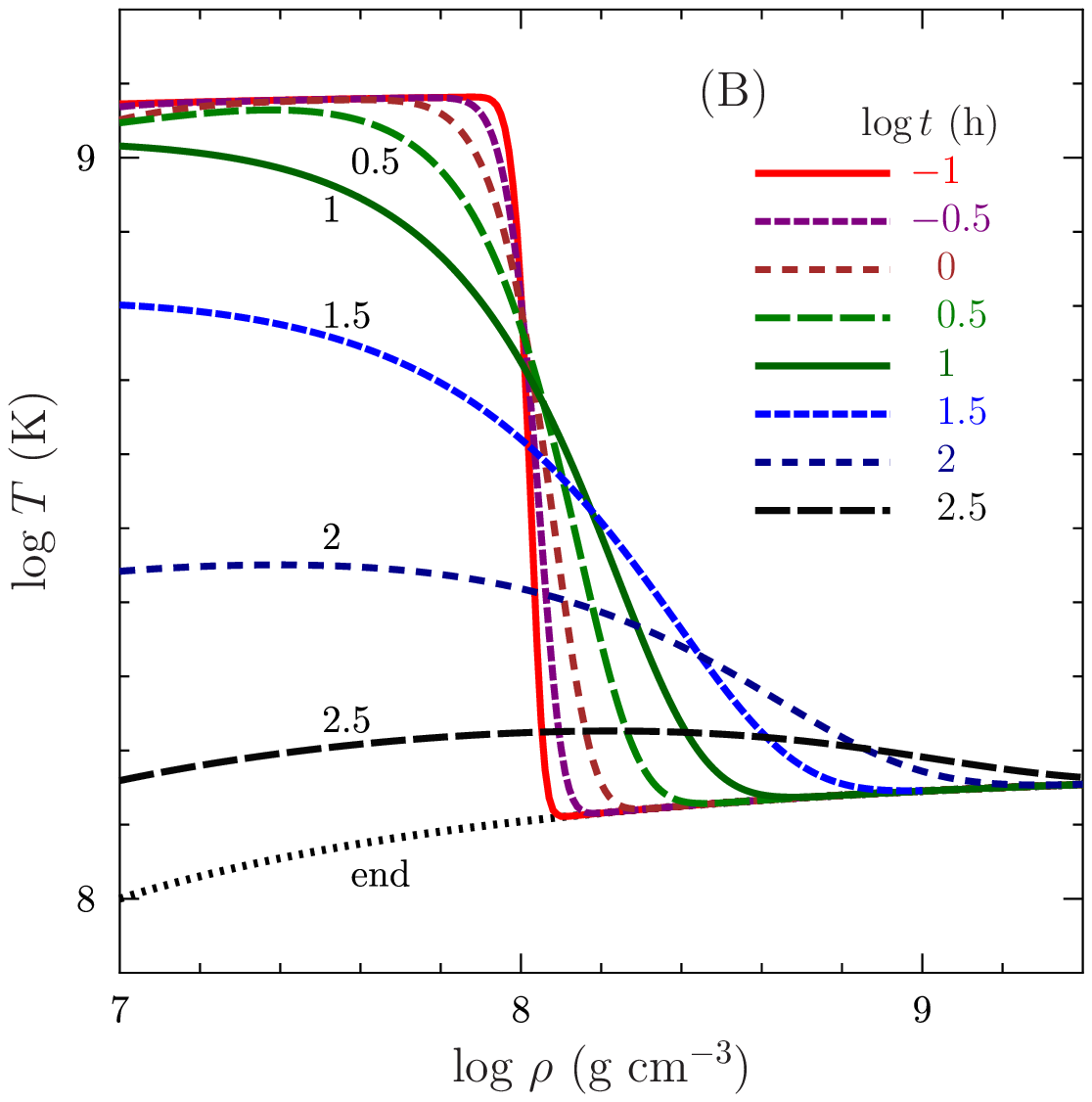}%
\hspace{5mm}
\includegraphics[width=0.44\textwidth]{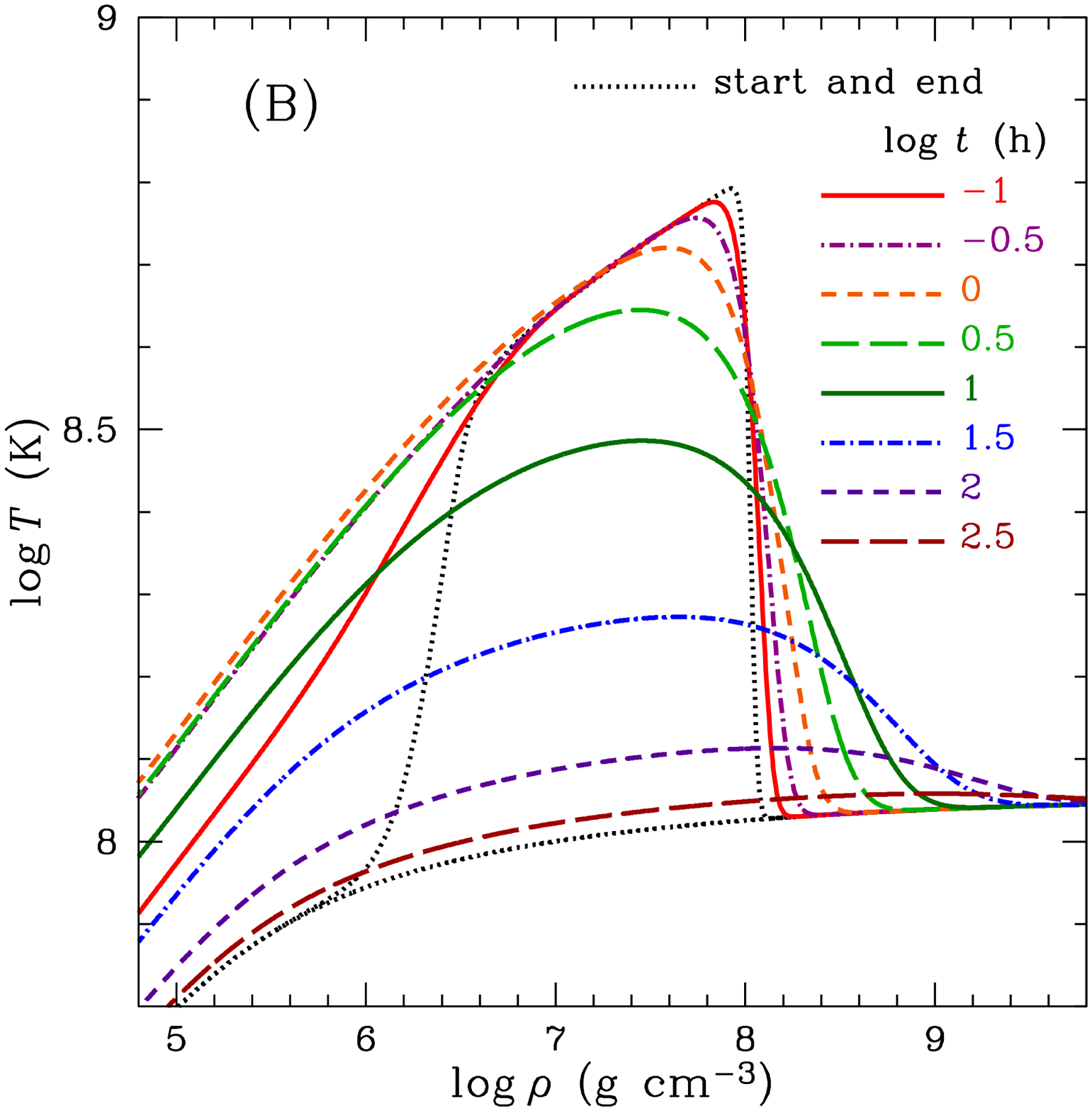}%
\caption{Same as in Fig.~\ref{f:TrhoAA} but for burst B.
}
\label{f:TrhoBB}
\end{figure*}

\begin{figure}
\includegraphics[width=0.45\textwidth]{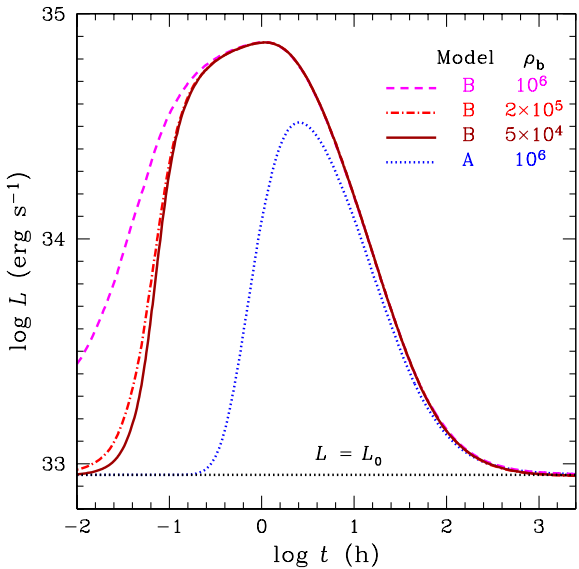}%
\caption{Surface luminosity versus time for bursts  from the four shells
extended from  
$\rho_\text{b}$   to $10^8$ \gcc\ (see the legend),
computed assuming a quasi-stationary outer envelopes of masses
$M_\text{env}=(10^{-12}$, $10^{-13}$, $10^{-14})$ {  \msun}, corresponding to
the $T$-dependent bottom densities $\rhob(T)$, whose 
approximate values $\sim(10^6$, $2\times 10^5$,  5$\times 10^4$) \gcc\
are marked in the legend. The dotted line for burst A (Fig.~\ref{f:LAA})
is shown for comparison with the other lines, which are computed for
model B. The lower horizontal  dotted line is the quiescent luminosity
(with $T_\text{0}=10^8$ K  at $\rho=10^7$ \gcc).   See the text for
details.
}
\label{f:LBB}
\end{figure}

Fig.~\ref{f:TrhoBB} shows snapshots of the temperature profiles
$T(\rho)$ in the outer crust of the star after   burst B (Table
\ref{tab:model}) at different moments of time $t$.  The curves are
calculated using the toy model (the left-hand panel) and the thermal
evolution code (the right-hand panel) under the assumption that $T=10^8$
K at $\rho=10^7$ \gcc{} prior to the burst. Fig.~\ref{f:TrhoBB} is
analogous to Fig.~\ref{f:TrhoAA} for burst A. As in
Fig.~\ref{f:TrhoAA}, the temperature profiles are traced to
lower $\rho$ in the right-hand panel.

Burst B is designed to be a more adequate representation of a realistic
superburst than burst~A. 
Let us recall that the toy model is justified at $\rho>\rhob=10^7$ \gcc,
while it is widely accepted that explosive carbon burning  in
superbursts occurs also at much lower densities, down to
$\rho_1 \sim 10^4$ \gcc\ or lower (e.g.
\citealt{2011KH,2012Keek}).  However, the main energy release takes
place  at $\rho \gtrsim \rhob$  so that the burning at lower densities
does not change the total energy budget, although it affects the
temperature distribution at $\rho \lesssim \rhob$, including densities
somewhat higher than $\rhob$.

To be consistent with standard
simulations of superbursts, in
the toy model B we use the solution of equation (\ref{e:Green1}), 
in which the heat source is extended to lower densities $\rho_{1}$, as in 
real superbursts. We have taken $\rho_1=3
\times 10^6$ \gcc; making $\rho_1$ still lower
would not change our results. This choice
of the toy-model solution gives realistic behaviour of the
temperature distribution $T(z,t)$ in the toy-model domain (\ref{e:rho,T}).
However, some extra energy is now released
in the density range $\rho < 10^7$ \gcc, where we use the heat-blanket
solution to calculate the lightcurve. The heat-blanket model is obtained 
without any additional short-term heating. Accordingly, we cannot rely on
our lightcurve as long as the extra heat is confined in the heat blanket
and the usual steady-state heat outflow is not established there.  

Using the above procedure, we obtain the toy-model
$T(\rho)$  profiles at $\rho \geq 10^7$ \gcc\  (the left-hand panel of
Fig.\ \ref{f:TrhoBB}) which are stable against  convection and resemble
those  obtained in advanced simulations of superbursts. 

The main difference of these profiles from those for toy burst A 
on the left-hand side of Fig.~\ref{f:TrhoAA} is the absence of 
temperature peaks associated with the
finite width of the heater A. The toy-model temperature 
gradient for burst B is mainly negative 
at $\rho \geq \rhob=10^7$ \gcc\ at all moments of time
because of the heat-{accumulation} nature of the toy model. In the density 
range $\rhob \leq \rho \lesssim \rho_2$, this gradient 
decreases with time, leading to the appearance of quasi-isothermal zones
(in $t \approx $ 30~h   for burst B).
The extra heat accumulated in this zone mainly sinks slowly 
inward
the star in the same manner as in burst A.   

The calculated toy-model temperature profiles are in reasonable
qualitative agreement with those computed using  the thermal evolution
code and presented on the right-hand panel of Fig.\ \ref{f:TrhoBB} (as
in Section \ref{s:burstAA} for burst A). 
However, the  toy model stronger overestimates $T$ at
$\rho \sim 10^7$ \gcc\ at earlier stages I and II, although the overall
agreement at the late stage III  is  satisfactory.  The slower toy-model
thermal diffusion is also quite visible. 
Apparently faster 
cooling of the heated layer in the numerical calculations
is realized because of stronger
heat outflow through the surface.
As explained above, the extra
energy, generated at $\rho < \rhob$, 
complicates construction of the 
toy-model lightcurve at the initial stages I and II of burst B, although
the agreement improves with time and becomes better at stage III.

In Fig.~\ref{f:LBB} we present three lightcurves calculated  by
the numerical code for burst B model ($\rho_1=3\times10^6$
\gcc) compared with one lightcurve for burst A 
($\rho_1=3\times10^7$ \gcc). The former three curves differ
by the
masses $M_\text{env}$ of  the 
quasi-stationary  envelope used in
simulations. We parametrize these masses by the approximate equivalent values of
 $\rhob$  listed in Section~\ref{s:check}.
Although the lightcurves for burst B are
somewhat different at stage I ($t \lesssim 10$ min), they merge in the single
curve later. At stage III  ($t \gtrsim 20$ h)
this curve is similar but slightly higher than the dotted
curve for burst A, because burst B is more energetic (Table
\ref{tab:model}). This similarity
is a genetic feature of lightcurves at stage III, as discussed below.
Very similar behaviour is due to the same bottom density $\rho_2$ of the
bursting shells in models B and A. 
According to the numerical simulations, about 25 per
cent of the energy released in burst B emerges through the surface. It
is higher than 20 per cent in burst A, because model B contains heating
layers located
closer to the surface.

\section{Generic features of bursts}
\label{s:generic}

Let us outline generic features of the stages I, II and III in the
evolution of deep bursts (Section \ref{s:stages}).

Stage I is short and dynamical. It can be accompanied or not accompanied
by convection, depending of the fuel distribution in the burning shell. 
For realistic bursts of thick shells filled with fuel to low densities
($\lesssim 10^4$ \gcc), convection seems unimportant
({because of reduction}  of temperature 
gradients).  
Stage
I ends with the onset of quasi-stationary flux-conserving heat outflow at 
$\rho \sim (10^6-10^7)$ \gcc.

The next stage II of the strongest energy release through the surface
{  is accompanied by} quasi-equilibration of the heat propagation though the entire
bursting shell (down to the ignition depth $\rho_2$). 
According to many simulations 
(e.g.\ \citealt{2004CM,2006Cumming,2012Altamirano}), 
the lightcurves $L(t)$ show a 
rapid initial rise 
(not always observable)
followed by a slow, e.g.\ power-law fall;
the power-law index is often treated as universal. 
Typically, the initial afterburst temperature profiles $T_\text{f}(\rho)$
gradually increase with density. 
{  However}, as demonstrated by \cite{2015Keek}, one can obtain steeper   
 profiles, for instance assuming that the fraction of burnt fuel 
increases with $\rho$ within the heated layer. In this case, 
the authors obtained 
the $L(t)$ curves containing smooth peaks at the most energetic
stage. Lightcurves of both types, with a slow $L(t)$ fall and 
with a 
preceding
peak, have been observed.  

We remark that microphysics in bursting sources can be different, for
example due to different ignition densities and temperatures 
(see Section \ref{s:latedecayrate}). Accordingly, 
we do not expect that the $L(t)$-profile at  burst stage II is universal. 
Varying microphysics and the fuel distribution, one can construct rather 
sophisticated profiles.

By the end  of 
the most energetic burst stage II,
the quasi-stationary regime    
 of flux-conserving  heat outflow 
(e.g. \citealt{2006Cumming})
is established
from 
the outer zone  
to the bottom of the initially heated layer,
$\rho_1 \leq \rho \lesssim  \rho_2$.

Recall that the temperature becomes almost independent of heat capacity and 
thermal conductivity in nearly
isothermal zones. Once such zones appear, calculated values
of $T$ within them start to be insensitive to the 
underlying microphysics.

Before the temperature equilibrates in the entire heated zone 
during stage II, the heat has not enough time 
to sink deeply inside the crust. The latter sinking mainly proceeds
at the final stage III of the burst.

\section{Late stage of burst decay}
\label{s:fading}

Here we focus on stage III of late burst
decay. So far our consideration was restricted to one neutron star model ($M=1.4\,\msun$,
$R=12$ km) and to fixed
ignition density ($\rho_2=10^8$ \gcc). 
We also restricted ourselves
to unrealistically low 
fuel calorimetry, in order to 
meet the assumptions inherent to the toy
model (in particular, the neglect of neutrino emission). 
In this section we will base on generic properties of bursts 
(Section~\ref{s:generic}) and perform 
a semi-quantitative analysis of the late decay stage III 
for rather arbitrary neutron
star models, ignition depths, and burst energies. 
Our consideration will also be independent
of possible strong neutrino cooling of the bursting shell at earlier
stages I and II. The analysis will be not too rigorous but hopefully reproduces
the main features of stage III under the assumption that the 
internal temperature is much higher than {  in quiescence}, $T \gg T_0$.

\subsection{Transition time to the late-decay stage}
\label{s:transit}

\begin{figure*}
\includegraphics[width=0.45\textwidth]{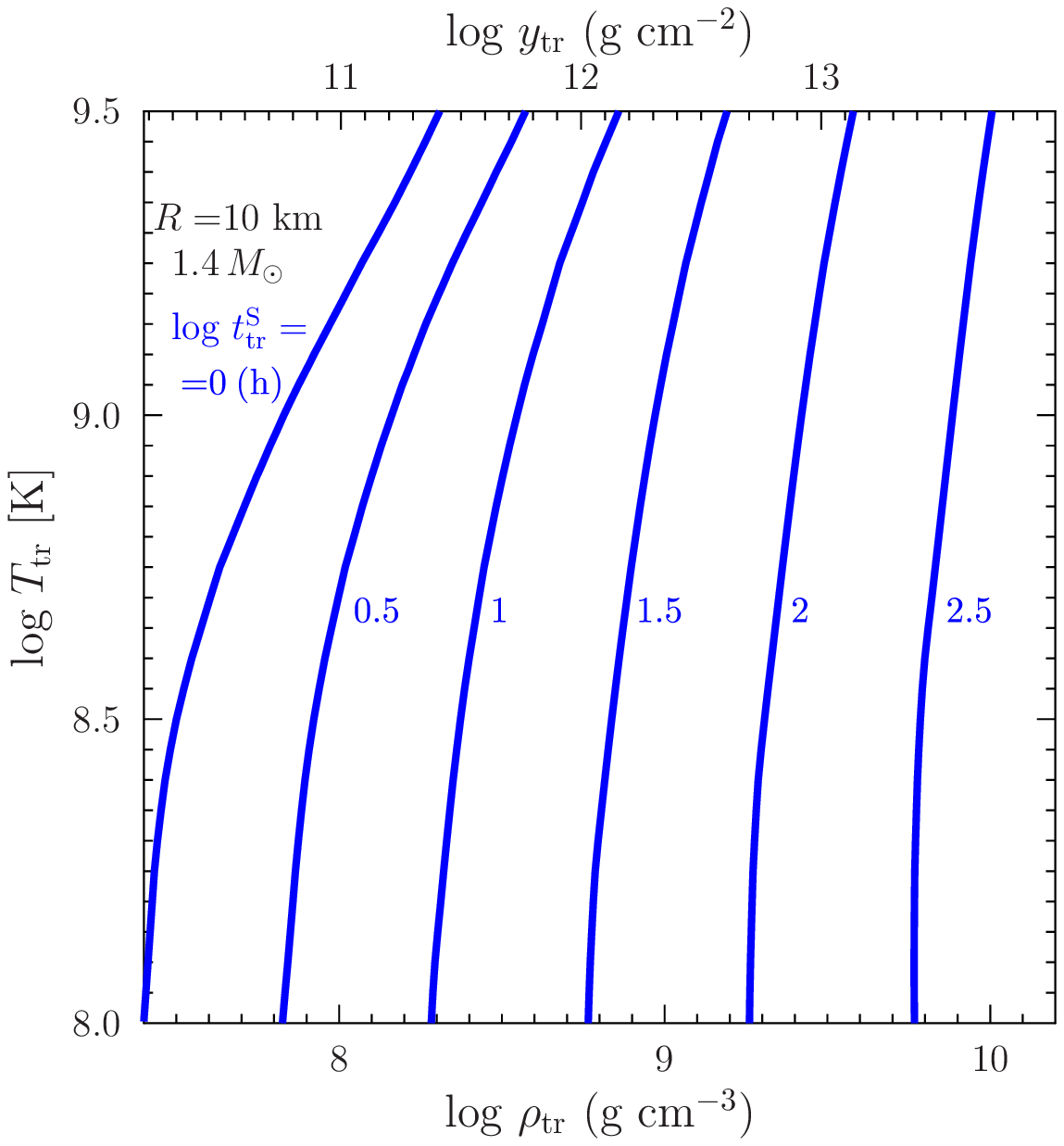}%
\hspace{5mm}
\includegraphics[width=0.45\textwidth]{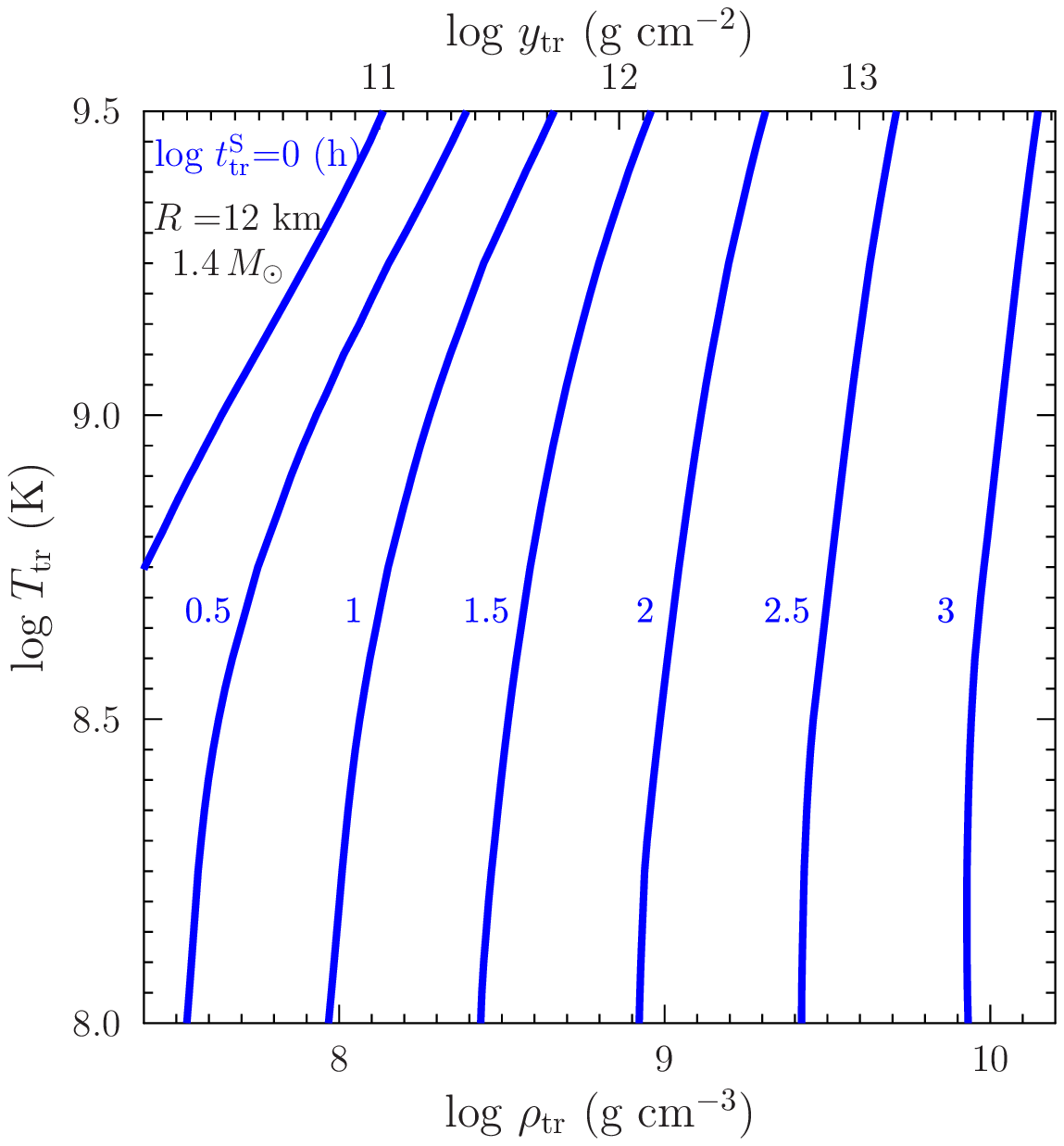}%
\caption{Isolines of constant thermal diffusion times 
($\log t^\text{S}_\text{tr}{\rm [h]}=$0, 0.5, 1, 1.5, 2, 2.5 and 3) 
in the $\log \rho_\text{tr}-\log T_\text{tr}$ plane for neutron stars 
with $M=$1.4 \msun\ and two values of $R$=10 km (left-hand panel) 
and 12 km (right-hand panel). 
Upper horizontal scales show column densities $y_\text{tr}$ (instead 
of $\rho_\text{tr}$) as commonly used in the literature.
See the text for details.
}
\label{f:timertr}
\end{figure*}

The transition from the most energetic stage II to the 
final stage III can be observable as a transition to faster 
lightcurve decay. Let $t_\text{tr}$ be the corresponding transition time
and $T_\text{tr}$ be the temperature in the nearly isothermal zone at
this epoch. It is natural to state  
(e.g.\ \citealt{2004CM,2006Cumming,2012Altamirano}) 
that $t_\text{tr}$ is the time of thermal wave propagation from the 
bottom of the heater through the entire outer zone (from the ignition
density $\rho_2 \approx \rho_\text{tr}$ to 
the surface). This time can be estimated as \citep{hl69}
\begin{equation}
        t_\text{tr}= \frac{1}{4}\, \left| \int_{0}^{z_2} 
        {\rm d}z \, \sqrt{\frac{C}{\kappa}} \, \right|^2,
    \label{e:tdiff}    
\end{equation}
where the integration is along the $T(z)$ track at $t\sim t_\text{tr}$.
For deep and strong bursts, the internal temperature  by that time
becomes nearly uniform, $T\approx T_\text{tr}$, over the most important
part of the track which contributes mainly to the integral.

In the toy model, it is sufficient to replace the lower integration
limit by $z_\text{b}$ (that is appropriate to $\rhob=10^7$ \gcc).
Using equations (\ref{e:C}) and (\ref{e:kappa}), we obtain
\begin{equation}
        t_\text{tr}=\frac{a}{{9}\  b} \, (z_2^{3/2}-z_\text{b}^{3/2})^2 
        \approx \frac{a}{{9}\ b} \, z_2^3 \approx   {0.35} 
             \,\rho_\text{tr6}~{\rm h};
    \label{e:toytr}    
\end{equation}
the final expression is obtained by setting
$z_\text{2} \gg z_\text{b}$, and the estimate 
is given for $M=1.4$~\msun\ and $R=12$ km, with
$\rho_\text{tr6}=\rho_\text{tr}/10^6$ \gcc. This gives  $t_\text{tr} \approx 35$~h  
 for toy bursts A and B.

\begin{figure}
\includegraphics[width=0.45\textwidth]{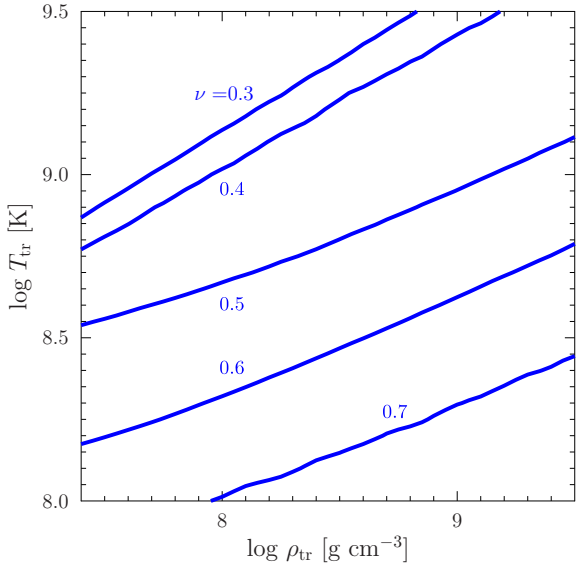}%
\caption{Isolines of constant power-law indices in equation
(\ref{e:general}), $\nu=$0.3, 0.4, 0.5, 0.6 and 0.7, on the $\log
\rho_\text{tr}-\log T_\text{tr}$ plane for neutron stars  (independent
of $M$ and $R$). See the text for details.
}
\label{f:pwindnu}
\end{figure}

Disregarding the toy model, we have calculated 
$t_\text{tr}$ from equation (\ref{e:tdiff}) 
along the $T=T_\text{tr}$ tracks for a dense 
grid of $\log \rho_\text{tr}~[ \gcc]$ 
(from 7.5 to 10 with step of 0.1)  
and $\log T_\text{tr}~[{\rm K}]$ 
(from 8 to 9.5 with step of 0.1). We 
have taken the lower integration limit at 
$\rho=10^6$ \gcc\  and used full realistic physics input. 
The calculated values present 
realistic estimates of thermal diffusion time 
from depths $\rho_\text{tr}$ at temperatures
$T_\text{tr}$ to the surface. The entire family 
of these diffusion times can be
fitted by 
\begin{equation}
t^{(0)}_\text{tr}[{\rm h}]= p_1 \rho_\text{tr6}^{p_3+1}/(1+p_2 \rho_\text{tr6}^{p_3}),
\label{e:fitt}
\end{equation}
where
\begin{eqnarray}
p_1&=&\frac{0.07483(l_T-7.786)}{1+\exp(11.37(l_T-8.576))}+3 \times 10^{-5},
\nonumber \\
p_2&=&\frac{0.5582(l_T-7.777)}{1+\exp(10.87(l_T-8.587))}+3.9 \times 10^{-4},
\nonumber \\
p_3&=& 0.8209+0.3865 \, \sin(3.658(l_T-14.04)),
\nonumber
\end{eqnarray}
and $l_T=\log_{10} (T_\text{tr}~[\rm K])$. 
The maximum relative fit error  is about 9 per cent (at $\log \rho_\text{tr}[\gcc]=7.5$ and
$\log T_\text{tr}[{\rm K}]=8$) and the rms relative error is 3 per cent, 
quite sufficient for our  semi-quantitative analysis.    
The superscript $(0)$ in  $t^{(0)}_\text{tr}$ indicates that the calculated values refer
to a star with $M=1.4$~\msun, $R=12$ km  and 
$g_\text{s0}=1.59 \times 10^{14}$
cm~s$^{-2}$.    
    
Note that 
we determine $t_\text{tr}$ 
in the local reference frame. According to 
equation (\ref{e:infty}) and
simple self-similarity arguments, 
a distant observer would measure 
\begin{equation} 
    t^\text{S}_\text{tr}= \,\frac{g_\text{s0}^2}{g_\text{s}^2}\,
    \frac{t_\text{tr}^{(0)}}{\sqrt{1-r_\text{g}/R}}
\label{e:ttrgen}    
\end{equation}
for a neutron star with arbitrary values of $M$ and $R$ (and corresponding
surface gravity $g_\text{s}$).

For example, Fig.\ \ref{f:timertr} shows the isolines of constant
$\log t_\text{tr}[{\rm h}]$=0.5, 1, 1.5 and 2 
in the $\rho_\text{tr}-T_\text{tr}$ plane for neutron stars with
the canonical mass $M=1.4$~\msun\ but two values of radius $R=10$ km (the
left-hand panel) and $R=12$ km (the right-hand panel). It is seen that
$t_\text{tr}$ depends mainly on $\rho_\text{tr}$. For the same $\rho_\text{tr}$,
the diffusion time is noticeably shorter in a more compact star
because its crust is geometrically thinner. In the toy model, the
diffusion time (\ref{e:toytr}) appears somewhat larger because 
our toy-model thermal conduction is slower. 

Inferring $t_\text{tr}$ from observations allows one to estimate
the ignition depth $\rho_\text{tr}$ (or ignition column 
$y_\text{tr}$) and put constraints on
possible mass and radius of the star. At $t \sim t_\text{tr}$ one
expects $T_\text{tr}\sim (1-3)\times 10^9$ K and the  luminosity of 
a real superburst 
$L_\text{tr}\sim (5-50)\times 10^{35}$ erg s$^{-1}$. 

\subsection{Late decay rate}
\label{s:latedecayrate}

Here we analyse the burst decay rate at $t \gtrsim t_\text{tr}$.

First of all we note that the transition temperature $T_\text{tr}$ can 
roughly be estimated as the temperature of the outer layer
$\rho_\text{min} \lesssim \rho \lesssim \rho_2$ heated by the nuclear 
column energy (that is, the energy  per unit surface area),
$H_\text{0tr}$, 
released
in the outer layer by 
the moment
$t \sim t_\text{tr}$. Generally, the initial heat $H_0$ can be 
transported outward (to
the surface), inward (to the core) and carried away  by neutrinos from the
heated layer.  The inward heat transport is slow; it can
usually be ignored at $t \lesssim t_\text{tr}$.  The heat transport to
the surface can be efficient at $t \lesssim t_\text{tr}$ but later it
becomes less important compared with the inward heat flux. The neutrino
emission (which we ignore in the bulk of this paper) can substantially
reduce the thermal  energy but this energy loss is quick (because it is
the strongly $T$-dependent, see \citealt{Yak2001}). It is expected to
become weak at $t \gtrsim t_\text{tr}$. Thus we can estimate
$T_\text{tr}$ as
\begin{equation}
     T_\text{tr} \sim H_\text{0tr}/(C_\text{tr}z_\text{tr}),
\label{e:Ttr}
\end{equation}
where $C_\text{tr}=C(z_\text{tr},T_\text{tr})$.

The last stage of the burst decay is controlled by 
sinking of the heat inside the star. 
The heat capacity of the outer crust under the heater is so
large that the matter easily absorbs the spreading heat. 
Accordingly, in spite of high thermal conductivity, 
the heat wave moves slowly inside the star, {  increasing
heat-accumulating property of the outer crust}.

This slow heat sinking is governed by equation (\ref{e:diff}) 
with $Q=0$. At $t \gtrsim t_\text{tr}$ an approximate solution can
be obtained using self-similarity properties inherent to this equation.
Let $z_*$, $T_*$, $C_*$ and $\kappa_*$
be, respectively, characteristic depth, excess temperature, 
heat capacity and thermal conductivity
of the inner front  
($z_* \gtrsim z_\text{tr} $)
of the spreading heat at moment $t$. 
At $t \gtrsim t_\text{tr}$ one can use the following
order-of-magnitude estimates,
\begin{equation}
T_* C_* z_* \sim H_\text{0tr}, \quad t \sim C_* z_*^2/\kappa_*.
\label{e:selfsim}
\end{equation}
The first estimate ensures approximate conservation of the
overall heat content  over time,   
in agreement with equation (\ref{e:Ttr}), and the second 
one describes ordinary heat diffusion.

Now let us assume arbitrary power-law dependences
of $C$ and $\kappa$ on temperature and density
\begin{subequations}
\label{e:Ckappaself}
\begin{eqnarray}
 C(z,T) & = & C_\text{tr} \left( \frac{z}{z_\text{tr}} \right)^{\alpha_1}
\left( \frac{T}{T_\text{tr}} \right)^{\alpha_2} ,
\\
 \kappa(z,T) & = & \kappa_\text{tr} \left( \frac{z}{z_\text{tr}} \right)^{\beta_1}
\left( \frac{T}{T_\text{tr}} \right)^{\beta_2},
\end{eqnarray}
\end{subequations}
and analyse the thermal 
wave propagation recalling that $\rho \propto z^3$,
in deep layers of the outer crust.
Here, $C_\text{tr}$ and $\kappa_\text{tr}$ normalize the heat capacity and thermal
conductivity, while $\alpha_1$, $\alpha_2$, $\beta_1$ and $\beta_2$ specify their 
density and temperature dependence. For a normalization point, we
take $z=z_\text{tr}$ at the moment $t=t_\text{tr}$ at which  the late afterburst
relaxation stage III starts, with $T_*=T_\text{tr}$. We substitute
equations (\ref{e:Ckappaself}) into (\ref{e:selfsim}) 
and obtain two equations containing powers of $t/t_\text{tr}$, $z_*/z_\text{tr}$
and $T_*/T_\text{tr}$. These equations yield the self-similar solution
for arbitrary microphysics of the matter in the form of power-law decay,
\begin{equation}
    \frac{T_* }{ T_\text{tr}} \sim
     \left( \frac{t_\text{tr}}{t} \right)^\nu
    \sim
     \left( \frac{z_\text{tr}}{z_*}  \right)^{(\alpha_1+1)/(\alpha_2+1)},
\label{e:general}
\end{equation}
where the power index is
\begin{equation}
   \nu = \frac{1+\alpha_1}
{2 + \alpha_1+ \alpha_2 
+ (1+\alpha_1) \beta_2 - (1+\alpha_2 ) \beta_1}.
\label{e:nu}
\end{equation}

The toy model corresponds to $\alpha_1=3$, $\beta_1=2$, $\alpha_2=\beta_2=0$
and gives $\nu=4/3$. At $H_\text{0tr} \sim H_0$ 
equation (\ref{e:general}) qualitatively reproduces the exact asymptotic
solution (\ref{e:Greenasy}) of the toy model problem for the internal temperature
decay and yields the law of
thermal wave propagation inside the star, 
$\rho_*/ \rho_\text{tr} \sim (z_*/z_\text{tr})^3 \sim t/t_\text{tr}$, which 
agrees with the toy model calculations. However, let us stress that, for
the parameters employed, the exact toy asymptotic regime is realized too late
to be observed in real events.

Nevertheless, the real plasma deviates from the toy-model due to  the complexity
of microphysics involved (Section \ref{s:toymicro}). We see 
that the index $\nu$  reflects
the rate of temperature decrease in the vicinity of the
depth $z_\text{tr}$ at the moment of time $t_\text{tr}$.
This $\nu$ is local (depends
on local density $\rho_\text{tr}$ and temperature $T_\text{tr}$),
 but independent of neutron
star mass and radius. For any pair of $\rho_\text{tr}$ and $T_\text{tr}$ we have
calculated the parameters $\alpha_{1,2}$ and $\beta_{1,2}$
in equation (\ref{e:Ckappaself}) as local power-laws, e.g.,
$\alpha_1=\partial \ln C(z,T)/\partial \ln z$ (at $z=z_\text{tr}$
and $T=T_\text{tr}$) with accurate microphysics of the matter.
Then we have found $\nu$ from equation (\ref{e:nu}). Fig.\ \ref{f:pwindnu}
presents isolines of constant $\nu$=0.3, 0.4, 0.5, 0.6 and 0.7
in the $\rho_\text{tr}-T_\text{tr}$ plane. One can see that higher $\nu$
are realized at sufficiently large densities and low temperatures.

Indeed, the microphysics of the matter varies with density and temperature.
For instance, at densities $\rho \sim 10^7$ \gcc\ the approximation
of temperature-independent thermal conductivity of degenerate 
electrons (\ref{e:kappa}) is violated,
and the linear dependence $\kappa \propto T$ becomes 
more suitable (e.g. \citealt{1999Palex}). Then,
as long as the degenerate electrons remain relativistic, 
one has $\kappa \propto T \rho^{1/3}$,
which corresponds to $\beta_1=\beta_2 \approx 1$. 
As for the heat capacity, it may 
still be mostly provided by ions, with $C \propto \rho$ 
($\alpha_1=3$, $\alpha_2=0$), as in the toy model;
see equation (\ref{e:C}). In that case, we
have $\nu=1/2$, 
so that $T_*/T_\text{tr} \sim \sqrt{t_\text{tr}/t}$ and 
$\rho_* / \rho_\text{tr} \sim (t/t_\text{tr})^{3/8}$.
Note that \cite{2004CM} obtained 
$\nu=2/3$ which is largely
used in the literature. Fig.\ \ref{f:pwindnu} demonstrates that
a wide range of $\nu$ can be realized in one star.

Now we can outline the behaviour of the lightcurve
$L(t)$ at the decay stage III. To this aim, we can assume that
at $t \gtrsim t_\text{tr}$ the internal temperature profile at densities
$ \rhob \lesssim \rho \lesssim \rho_*$ is nearly flat, 
$T(\rho) \approx T_*$. Then
we can estimate the surface temperature $T_\text{s}$ 
and the surface luminosity
$L(t)=4 \pi R^2 \sigma_\text{SB} T_\text{s}^4$ with the aid of
the $T_\text{s}-T_\text{b}$ relation for the heat
blanketing envelope (Section \ref{s:heat-blanket}). These results can be 
roughly approximated by the power-law decay
\begin{equation}
L(t)\approx (t_\text{tr}/t)^\gamma \, L_\text{tr},
\label{e:Ldecay} 
\end{equation} 
with the index 
$\gamma={\rm d} \ln L(t)/{\rm d} \ln t$ at $t=t_\text{tr}$.
Unfortunately, our calculations show that accurate values
of $\gamma$ do depend on deviations of $T(\rho)$ from
constants $T_\text{tr}$ near the heat blanket 
($\rho > 10^7$ \gcc) and on exact behaviour 
of the $T_\text{s}-T_\text{b}$ relation at $\rho=\rho_\text{b}$.
The problem of accurate calculation of $\gamma$ deserves a
special study. The robust conclusion is that
$\gamma$ is larger for deeper bursts (higher $\rho_\text{tr}$) and
lower $T_\text{tr}$. These results do not support the
idea that $\gamma$ is universal for all superbursts.

One should bear in mind that the self-similar 
approach is only an approximation {  based on the assumptions 
that local miscophysical parameters, such as $\alpha_{1,2}$ and
$\beta_{1,2}$ vary sufficiently slowly}. It is natural that as the crust is cooling at stage III, 
microphysics of 
the characteristic density and temperature domain, that controls the
cooling, is changing. This may lead to a variable $\gamma$ along the
cooling track.    

\subsection{Analysing late superburst tails}
\label{s:analyze}

The results on times $t \gtrsim t_\text{tr}$ of late superburst decay onset
(Section \ref{s:transit}) and on lightcurve slope $\gamma$ during 
the late tail stage III (Section \ref{s:latedecayrate})
can be used for a preliminary semi-quantitative `express' analysis of superbursts.
A transition time $t^\text{S}_\text{tr}$  from stage 
II to stage III can be observed as
a change of the lightcurve slope (from a slow to faster decay).
A power-law index $\gamma$ can potentially be inferred from an
observed lightcurve at stage III. Note that both  measurements 
(of $t^\text{S}_\text{tr}$ and $\gamma$) do
not require normalization of  lightcurves. It is worth
to remark that the accurate determination
of $\gamma$ is a serious problem because of large errorbars of $L(t)$ 
at the tail stage when the source is fading.  
Analysing $t_\text{tr}$ seems more informative.

For instance, let us consider six superbursts detected
with {\it BeppoSAX} and analysed
by \citet{2006Cumming}. They were the superbursts from 4U 1524--690 
observed in 1999 \citep{2003Zand};  4U 1735--444 
(1996, \citealt{2000Cornelisse}); KS 1731--260
(1996, \citealt{2002Kuulkers}); GX 17+2 (1999, \citealt*{2004Zand}),
Ser X-1 (1997, \citealt{2002Cornelisse}); 4U 1636--536
(2001, \citealt{2002SM,2004Kuulkers}). The observed lightcurves and
theoretical fits are given in figs. 5--10 of \citet{2006Cumming};
the fit parameters are listed in table 1 of that paper. 

Let us take, for instance, the KS 1731--260 superburst (fig.\ 5) and
determine $t_\text{tr}^\text{S}$ as the time after which the theoretical
fit becomes nearly power-law. We have  $t_\text{tr}^\text{S}\approx 10$ h.
Since \citet{2006Cumming} took $M=1.4\,\msun$ and $R=10$ km for their
interpretation, we use the left-hand panel of Fig.\ \ref{f:timertr},
adopt $\log T_\text{tr}$ [K]$\approx 9-9.3$ and obtain 
$\log y_\text{tr}~[{\rm g~cm^{-2}}]\approx 12$ and 
$\log \rho_\text{tr} [\gcc] \approx 8.7$, in nice agreement 
with \citet{2006Cumming}. Similar agreement takes place for other
five superbursts.  Note, however, that in order to explain the
reported $\log  y_\text{tr}~[{\rm g~cm^{-2}}]\approx  11.7 $ 
($\log \rho_\text{tr} [\gcc] \approx   8.5 $)
for the shortest superburst ($t_\text{tr}^\text{S} \approx 2$ h), 
demonstrated by 4U 1636--536, we need to assume higher 
$\log T_\text{tr}~[{\rm K}]\approx 9.5$. 

Now let us return to the KS 1731--260 superburst {  and} take the same
$M=1.4\, \msun$ but larger $R=12$ km. In this case, we should
use the right-hand panel of Fig.\ \ref{f:timertr}. With the same
$t_\text{tr}^\text{S}$ and $\log T_\text{tr}$ we obtain 
$\log y_\text{tr}[\mbox{g cm$^{-2}$}] \approx 11.3$ and 
$\log \rho_\text{tr}[\gcc]\approx 8.2$. With the larger radius $R$, 
the crust becomes thicker, which makes thermal diffusion slower. 
Accordingly the ignition density
$\rho_\text{tr}$ has to be about three times smaller to ensure
the same time  $t_\text{tr}^\text{S}$ for the late stage III
onset. Similar shifts of the ignition density to the surface would
take place for other superbursts, meaning that theoretical interpretation
of superbursts is rather sensitive to neutron star mass and radius. 
Our Fig.\ \ref{f:timertr} and equations (\ref{e:fitt})
and (\ref{e:ttrgen}) can be helpful for understanding which 
$M$ and $R$ are more suitable.

Let us mention again the remarkable superburst of 4U 1636--536 with
$t_\text{tr}^\text{S}\approx 2$ h. Recall that \citet{2006Cumming} 
assumed $M=1.4\,\msun$ and $R=10$ km and obtained 
$\log y_\text{tr}~[\mbox{g cm$^{-2}$}] \approx 11.7$ and
$\log \rho_\text{tr}~[\gcc]\approx 8.5$. \citet{2015Keek}
adopted the same $M$ but $R=12$ km and obtained 
$\log y_\text{tr}~[\mbox{g cm$^{-2}$}] \approx 11.3$ and
$\log \rho_\text{tr}~[\gcc]\approx 8.2$, in agreement with
the right-hand side of Fig.\ \ref{f:timertr}. The interesting
feature of this source is that the superburst tail has been
measured to rather low luminosities 
(fig. 2 in \citealt{2015Keek}). Although the tail measurements 
show substantial time variations, they might be interpreted in
a way that the late tail decays faster (with larger $\gamma$)
than its beginning. \citet{2015Keek} attribute this effect
to some instabilities in the accretion disc. 

We {  would like to} note that there may be another explanation associated with
the genuine acceleration of the crustal cooling at stage III. 
According to Fig.\ \ref{f:pwindnu},
as the temperature goes down in the crust at the tail stage, the local
power-law 
$\nu$ can substantially increase and accelerate the cooling 
(Sect. \ref{s:latedecayrate}). 
Note that 
the neutron star can be more compact (for instance, 
it could be more massive). 
Then the ignition is shifted to higher  densities, 
which facilitates the process.  
This is just a possibility which might 
be checked in detailed simulations.

\section{Conclusions}
\label{s:conclude}

We have developed a simplified analytic model (`toy model')
to study heat diffusion  
after a burst in deep layers of the outer neutron star crust,
at sufficiently high densities and temperatures, see equation~(\ref{e:rho,T}). 

The applicability of this model is quite restricted.
It cannot follow nuclear reaction networks and associated
evolution of microphysical properties of the matter. It does not allow one to 
study the stages of accretion, accumulation and procession of nuclear fuel, 
the appearance of shocks and precursors before a burst, 
dynamics of nuclear burning and nucleosynthesis,
heat outflow due to neutrino emission 
(in contrast to modern computer codes,
e.g. \citealt{2006Cumming,2011KH,2012Keek,2015Keek,2017Galloway,2017Zand}
and references therein).

However, the toy model is simple and requires no special computer resources.
It can simulate important fragments of real events 
and predicts generic features of real bursts.

It is important that a warm outer crust of a neutron star
has large heat capacity and operates as a huge heat reservoir. It can easily
keep the heat generated in a burst for a few   months. 
Generic features include the appearance of a
quasi-isothermal zone
above the layer, where the main burst energy is released
and a very slow heat diffusion to the inner crust. This leaves
the bottom of the outer crust sufficiently cold and thermally decoupled
from 
the heated zone in the upper layers.
The burst energy is mainly 
transported inside the star although some fraction can be carried away by 
neutrinos from the bursting layer while another fraction diffuses to the
surface and can be observable. Typically, the burst
 that is seen from the surface
fades before the heat wave reaches the inner crust. 
We have shown that the toy model can be useful to describe the
late stage of the afterburst relaxation.

Note that our method can be inaccurate at lower temperatures, $T
\lesssim  10^8$~K.  In that case, the heat capacity is strongly reduced
by quantum effects in the motion of ions. Also, the thermal conductivity
of degenerate electrons becomes essentially dependent on temperature and
on the presence of impurities (ions of different types; e.g.
\citealt{2015PPP}).  Moreover, the toy model cannot be directly applied
to the inner crust of the neutron star, where free neutrons appear in
the matter, in addition to atomic nuclei and strongly degenerate
electrons. These free neutrons are numerous there. If they were normal,
they would be the source of large heat capacity, but they most likely
are superfluid. Their superfluidity greatly reduces the heat capacity of
the inner crust (e.g. \citealt{2006Cumming}). Generally, the inner crust
seems to be a poorer heat reservoir than the outer crust (e.g.,
\citealt{HPY2007}).

The toy model can be generalized to include the effects of neutrino cooling. 
Such models can be used to guide more elaborated numerical simulations of 
bursting neutron stars.

\section*{Acknowledgments}

We are grateful to A.~I. Chugunov, M.~E. Gusakov, D.~D. Ofengeim  and
P.~S. Shternin for fruitful comments.   The work by DY and AP was
supported by the Russian Science Foundation (grant 19-12-00133). The
work by PH was partially supported by the National Science Center,
Poland  (grant 2018/29/B/ST9/02013). AK is grateful for excellent 
working conditions during his visit of Copernicus Astronomical Center in
Warsaw.

\section*{Data availability}
The data underlying this article will be shared on
reasonable request to the corresponding author.

\bibliographystyle{mnras}


\begin{thebibliography}{}
	\makeatletter
	\relax
	\def\mn@urlcharsother{\let\do\@makeother \do\$\do\&\do\#\do\^\do\_\do\%\do\~}
	\def\mn@doi{\begingroup\mn@urlcharsother \@ifnextchar [ {\mn@doi@}
		{\mn@doi@[]}}
	\def\mn@doi@[#1]#2{\def\@tempa{#1}\ifx\@tempa\@empty \href
		{http://dx.doi.org/#2} {doi:#2}\else \href {http://dx.doi.org/#2} {#1}\fi
		\endgroup}
	\def\mn@eprint#1#2{\mn@eprint@#1:#2::\@nil}
	\def\mn@eprint@arXiv#1{\href {http://arxiv.org/abs/#1} {{\tt arXiv:#1}}}
	\def\mn@eprint@dblp#1{\href {http://dblp.uni-trier.de/rec/bibtex/#1.xml}
		{dblp:#1}}
	\def\mn@eprint@#1:#2:#3:#4\@nil{\def\@tempa {#1}\def\@tempb {#2}\def\@tempc
		{#3}\ifx \@tempc \@empty \let \@tempc \@tempb \let \@tempb \@tempa \fi \ifx
		\@tempb \@empty \def\@tempb {arXiv}\fi \@ifundefined
		{mn@eprint@\@tempb}{\@tempb:\@tempc}{\expandafter \expandafter \csname
			mn@eprint@\@tempb\endcsname \expandafter{\@tempc}}}
	
	\bibitem[\protect\citeauthoryear{{Altamirano} et~al.,}{{Altamirano}
		et~al.}{2012}]{2012Altamirano}
	{Altamirano} D.,  et~al., 2012, \mnras, 426, 927
	
	\bibitem[\protect\citeauthoryear{{Baiko}, {Kaminker}, {Potekhin}  \&
		{Yakovlev}}{{Baiko} et~al.}{1998}]{BKPY}
	{Baiko} D.~A.,  {Kaminker} A.~D.,  {Potekhin} A.~Y.,   {Yakovlev} D.~G.,  1998,
	\mn@doi [\prl] {10.1103/PhysRevLett.81.5556}, \href
	{https://ui.adsabs.harvard.edu/abs/1998PhRvL..81.5556B} {81, 5556}
	
	\bibitem[\protect\citeauthoryear{{Bateman} \& {Erd\'elyi}}{{Bateman} \&
		{Erd\'elyi}}{1953}]{1953BE}
	{Bateman} H.,  {Erd\'elyi} A.,  1953, {Higher Transcendental Functions}.
	McGraw-Hill, New York
	
	\bibitem[\protect\citeauthoryear{{Chaikin}, {Kaminker}  \&
		{Yakovlev}}{{Chaikin} et~al.}{2018}]{2018Chaikin}
	{Chaikin} E.~A.,  {Kaminker} A.~D.,   {Yakovlev} D.~G.,  2018, \apss, 363, 209
	
	\bibitem[\protect\citeauthoryear{{Cornelisse}, {Heise}, {Kuulkers}, {Verbunt}
		\& {in~'t Zand}}{{Cornelisse} et~al.}{2000}]{2000Cornelisse}
	{Cornelisse} R.,  {Heise} J.,  {Kuulkers} E.,  {Verbunt} F.,   {in~'t Zand}
	J.~J.~M.,  2000, \aap, 357, L21
	
	\bibitem[\protect\citeauthoryear{{Cornelisse}, {Kuulkers}, {in~'t Zand},
		{Verbunt}  \& {Heise}}{{Cornelisse} et~al.}{2002}]{2002Cornelisse}
	{Cornelisse} R.,  {Kuulkers} E.,  {in~'t Zand} J.~J.~M.,  {Verbunt} F.,
	{Heise} J.,  2002, \mn@doi [\aap] {10.1051/0004-6361:20011591}, 382, 174
	
	\bibitem[\protect\citeauthoryear{{Cumming} \& {Macbeth}}{{Cumming} \&
		{Macbeth}}{2004}]{2004CM}
	{Cumming} A.,  {Macbeth} J.,  2004, \apj, 603, L37
	
	\bibitem[\protect\citeauthoryear{{Cumming}, {Macbeth}, {in~'t Zand}  \&
		{Page}}{{Cumming} et~al.}{2006}]{2006Cumming}
	{Cumming} A.,  {Macbeth} J.,  {in~'t Zand} J.~J.~M.,   {Page} D.,  2006, \apj,
	646, 429
	
	\bibitem[\protect\citeauthoryear{{Eichler} \& {Cheng}}{{Eichler} \&
		{Cheng}}{1989}]{ec89}
	{Eichler} D.,  {Cheng} A.~F.,  1989, \mn@doi [\apj] {10.1086/167015}, \href
	{https://ui.adsabs.harvard.edu/abs/1989ApJ...336..360E} {336, 360}
	
	\bibitem[\protect\citeauthoryear{{Galloway} \& {Keek}}{{Galloway} \&
		{Keek}}{2017}]{2017Galloway}
	{Galloway} D.~K.,  {Keek} L.,  2017, e-print arXiv:1712.0627 (in
	Belloni T., Mendez M., Zang C., eds, 2021, Neutron Stars:
	Pulsations, Oscillations, Explosions, Springer, Berlin,  p. 209) 
	
	\bibitem[\protect\citeauthoryear{{Gradshteyn} \& {Ryzhik}}{{Gradshteyn} \&
		{Ryzhik}}{2007}]{2007GR}
	{Gradshteyn} I.~S.,  {Ryzhik} I.~M.,  2007, {Table of Integrals, Series, and
		Products, Seventh Edition}.
	Elsevier, Amsterdam
	
	\bibitem[\protect\citeauthoryear{{Gudmundsson}, {Pethick}  \&
		{Epstein}}{{Gudmundsson} et~al.}{1983}]{1983GPE}
	{Gudmundsson} E.~H.,  {Pethick} C.~J.,   {Epstein} R.~I.,  1983, Astrophys. J.,
	272, 286
	
	\bibitem[\protect\citeauthoryear{{Haensel}, {Potekhin}  \&
		{Yakovlev}}{{Haensel} et~al.}{2007}]{HPY2007}
	{Haensel} P.,  {Potekhin} A.~Y.,   {Yakovlev} D.~G.,  2007, {Neutron Stars. 1.
		Equation of State and Structure}.
	Springer, New York
	
	\bibitem[\protect\citeauthoryear{{Henyey} \& {L'Ecuyer}}{{Henyey} \&
		{L'Ecuyer}}{1969}]{hl69}
	{Henyey} L.,  {L'Ecuyer} J.,  1969, \apj, \href
	{http://adsabs.harvard.edu/abs/1969ApJ...156..549H} {156, 549}
	
		\bibitem[\protect\citeauthoryear{{in~'t Zand}}{{in~'t Zand}}{2017}]{2017Zand}
	{in~'t Zand} J.,  2017, in {Serino} M.,  {Shidatsu} M.,  {Iwakiri} W.,
	{Mihara} T.,  eds, 7 years of MAXI: monitoring X-ray Transients. RIKEN, Wako, p.~121
	
	\bibitem[\protect\citeauthoryear{{in~'t Zand}, {Kuulkers}, {Verbunt}, {Heise}
		\& {Cornelisse}}{{in~'t Zand} et~al.}{2003}]{2003Zand}
	{in~'t Zand} J.~J.~M.,  {Kuulkers} E.,  {Verbunt} F.,  {Heise} J.,
	{Cornelisse} R.,  2003, \mn@doi [\aap] {10.1051/0004-6361:20031586}, 411,
	L487
	
	\bibitem[\protect\citeauthoryear{{in~'t Zand}, {Cornelisse}  \&
		{Cumming}}{{in~'t Zand} et~al.}{2004}]{2004Zand}
	{in~'t Zand} J.~J.~M.,  {Cornelisse} R.,   {Cumming} A.,  2004, \mn@doi [\aap]
	{10.1051/0004-6361:20040522}, 426, 257
	
	\bibitem[\protect\citeauthoryear{{Kaminker}, {Kaurov}, {Potekhin}  \&
		{Yakovlev}}{{Kaminker} et~al.}{2014}]{2014KAM}
	{Kaminker} A.~D.,  {Kaurov} A.~A.,  {Potekhin} A.~Y.,   {Yakovlev} D.~G.,
	2014, \mnras, 442, 3484
	
	\bibitem[\protect\citeauthoryear{{Keek} \& {Heger}}{{Keek} \&
		{Heger}}{2011}]{2011KH}
	{Keek} L.,  {Heger} A.,  2011, \apj, 743, 189
	
	\bibitem[\protect\citeauthoryear{{Keek}, {Heger}  \& {in~'t Zand}}{{Keek}
		et~al.}{2012}]{2012Keek}
	{Keek} L.,  {Heger} A.,   {in~'t Zand} J.~J.~M.,  2012, \mn@doi [ApJ]
	{10.1088/0004-637X/752/2/150}, \href
	{https://ui.adsabs.harvard.edu/abs/2012ApJ...752..150K} {752, 150}
	
	\bibitem[\protect\citeauthoryear{{Keek}, {Cumming}, {Wolf}, {Ballantyne},
		{Suleimanov}, {Kuulkers}  \& {Strohmayer}}{{Keek} et~al.}{2015}]{2015Keek}
	{Keek} L.,  {Cumming} A.,  {Wolf} Z.,  {Ballantyne} D.~R.,  {Suleimanov} V.~F.,
	{Kuulkers} E.,   {Strohmayer} T.~E.,  2015, \mn@doi [MNRAS]
	{10.1093/mnras/stv2124}, \href
	{https://ui.adsabs.harvard.edu/abs/2015MNRAS.454.3559K} {454, 3559}
	
	\bibitem[\protect\citeauthoryear{{Kuulkers} et~al.,}{{Kuulkers}
		et~al.}{2002}]{2002Kuulkers}
	{Kuulkers} E.,  et~al., 2002, \mn@doi [\aap] {10.1051/0004-6361:20011654}, 382,
	503
	
	\bibitem[\protect\citeauthoryear{{Kuulkers}, {in~'t Zand}, {Homan}, {van
			Straaten}, {Altamirano}  \& {van der Klis}}{{Kuulkers}
		et~al.}{2004}]{2004Kuulkers}
	{Kuulkers} E.,  {in~'t Zand} J.,  {Homan} J.,  {van Straaten} S.,  {Altamirano}
	D.,   {van der Klis} M.,  2004, in {Kaaret} P.,  {Lamb} F.~K.,   {Swank}
	J.~H.,  eds,  American Institute of Physics Conference Series Vol. 714, X-ray
	Timing 2003: Rossi and Beyond. pp 257--260 (\mn@eprint {arXiv}
	{astro-ph/0402076}), \mn@doi{10.1063/1.1781037}
	
	\bibitem[\protect\citeauthoryear{{Misner}, {Thorne}  \& {Wheeler}}{{Misner}
		et~al.}{1973}]{MisnerTW}
	{Misner} C.~W.,  {Thorne} K.~S.,   {Wheeler} J.~A.,  1973, {Gravitation}.
	W.~H. Freeman and Co., San Francisco
	
	\bibitem[\protect\citeauthoryear{{Pearson}, {Chamel}, {Potekhin}, {Fantina},
		{Ducoin}, {Dutta}  \& {Goriely}}{{Pearson} et~al.}{2018}]{Pearson_18}
	{Pearson} J.~M.,  {Chamel} N.,  {Potekhin} A.~Y.,  {Fantina} A.~F.,  {Ducoin}
	C.,  {Dutta} A.~K.,   {Goriely} S.,  2018, \mnras, \href
	{https://ui.adsabs.harvard.edu/abs/2018MNRAS.481.2994P} {481, 2994}
	
	\bibitem[\protect\citeauthoryear{{Potekhin} \& {Chabrier}}{{Potekhin} \&
		{Chabrier}}{2018}]{PC18}
	{Potekhin} A.~Y.,  {Chabrier} G.,  2018, \aap, 609, A74
	
	\bibitem[\protect\citeauthoryear{{Potekhin}, {Chabrier}  \&
		{Yakovlev}}{{Potekhin} et~al.}{1997}]{1997PCY}
	{Potekhin} A.~Y.,  {Chabrier} G.,   {Yakovlev} D.~G.,  1997, Astron.
	Astrophys., 323, 415
	
	\bibitem[\protect\citeauthoryear{{Potekhin}, {Baiko}, {Haensel}  \&
		{Yakovlev}}{{Potekhin} et~al.}{1999}]{1999Palex}
	{Potekhin} A.~Y.,  {Baiko} D.~A.,  {Haensel} P.,   {Yakovlev} D.~G.,  1999,
	\aap, 346, 345
	
	\bibitem[\protect\citeauthoryear{{Potekhin}, {Pons}  \& {Page}}{{Potekhin}
		et~al.}{2015}]{2015PPP}
	{Potekhin} A.~Y.,  {Pons} J.~A.,   {Page} D.,  2015, \ssr, 191, 239
	
	\bibitem[\protect\citeauthoryear{{Richardson}, {Savedoff}  \& {Van
			Horn}}{{Richardson} et~al.}{1979}]{Richardson_79}
	{Richardson} M.~B.,  {Savedoff} M.~P.,   {Van Horn} H.~M.,  1979, \apjs, \href
	{https://ui.adsabs.harvard.edu/abs/1979ApJS...39...29R} {39, 29}
	
	\bibitem[\protect\citeauthoryear{{Salpeter}}{{Salpeter}}{1961}]{Salpeter61}
	{Salpeter} E.~E.,  1961, \mn@doi [\apj] {10.1086/147194}, \href
	{https://ui.adsabs.harvard.edu/abs/1961ApJ...134..669S} {134, 669}
	
	\bibitem[\protect\citeauthoryear{{Strohmayer} \& {Markwardt}}{{Strohmayer} \&
		{Markwardt}}{2002}]{2002SM}
	{Strohmayer} T.~E.,  {Markwardt} C.~B.,  2002, \mn@doi [\apj] {10.1086/342152},
	577, 337
	
	\bibitem[\protect\citeauthoryear{{Yakovlev} \& {Urpin}}{{Yakovlev} \&
		{Urpin}}{1980}]{1980YU}
	{Yakovlev} D.~G.,  {Urpin} V.~A.,  1980, \sovast, 24, 303
	
	\bibitem[\protect\citeauthoryear{{Yakovlev}, {Kaminker}, {Gnedin}  \&
		{Haensel}}{{Yakovlev} et~al.}{2001}]{Yak2001}
	{Yakovlev} D.~G.,  {Kaminker} A.~D.,  {Gnedin} O.~Y.,   {Haensel} P.,  2001,
	Phys. Rep., 354, 1
	
	\bibitem[\protect\citeauthoryear{{Ziman}}{{Ziman}}{1960}]{ZIMAN}
	{Ziman} J.~M.,  1960, {Electrons and phonons}.
	Clarendon Press, Oxford
	

	
	\makeatother
\end{thebibliography}

\appendix

\section{Green's function}
\label{s:green}

We need to solve equation (\ref{e:diff1}) which is obtained from
equation (\ref{e:diff}) with $C=a z^3$ and $\kappa=b z^2$ in accordance
with (\ref{e:C}) and (\ref{e:kappa}). Instead, we will be 
more general here and set 
\begin{equation}
    C=a z^\alpha, \quad \kappa=b z^\beta,
    \label{a:Ckappa}
\end{equation}
with arbitrary $\alpha$ and $\beta$, {  assuming constant values of} $a$
[$\text{erg}\, \text{cm}^{-\alpha-3}\,\text{K}^{-1}$]
 and $b$
[$\text{erg}\,\text{cm}^{-\beta-1}\,\text{s}^{-1}\,\text{cm}^{-1}$].
Then the equation to be solved reduces to
\begin{equation}
a z^\alpha {\partial \over \partial t}\ T - b\, {\partial \over \partial z}
\left( z^\beta {\partial \over \partial z}\ T  \right) = Q(z,t). 
\label{a:diff1}
\end{equation}

Let us use the Laplace transformation of equation (\ref{a:diff1}) with respect to
$t$ at
$Q=0$,
\begin{equation}
\widetilde{T}(z, s)= \int_0^\infty \dd t\ \exp(-s t) \, T (z, t), 
\label{Laplace}
\end{equation}
and introduce a dimensionless variable $x$,  
\begin{equation}
x= \left( {z \over z_s} \right)^{\mu} = u \sqrt{s},\quad
u=\frac{z^\mu}{\mu}\,\sqrt{\frac{a}{b}},
\label{u}
\end{equation}
with $\mu=(\alpha-\beta+2)/2$.
Then we obtain
the second-order  homogeneous  differential equation 
\begin{equation}
x^2 \widetilde{T}'' + \frac{\alpha+\beta} {\alpha-\beta+2} \, x   
\widetilde{T}'
- x^2\,\widetilde{T} = 0 
\label{dTdu}
\end{equation}
for $\widetilde{T}$ as a function of $x$; primes denote differentiation over
$x$.

Introducing $\lambda=(\beta-1)/(\alpha -\beta+2)$ and 
$Y=\widetilde{T}\,x^\lambda$,
we come to the Bessel equation of imaginary argument
\begin{equation}
x^2  Y'' +  x  Y'
- \left( x^2 + \lambda^2 \right) Y = 0, 
\label{Bessel}
\end{equation}
for $Y=Y(x,s)$ as a function of $x$.
A general solution of equation (\ref{dTdu}) for 
$\widetilde{T}$ is  
\begin{equation}
\widetilde{T} (x,s) = {1 \over x^{\lambda}} 
\left[ D_1(s)  K_{\lambda}(x) + D_2(s) I_{\lambda}(x) \right], 
\label{GenT}
\end{equation}
where 
$I_{\lambda} (x)$ and $K_{\lambda} (x)$ are the modified Bessel functions;
$D_1(s)$ and $D_2(s)$ remain to be determined.  

The Laplace transform of  
the thermal flux density  
$q (z, t) = - \kappa \partial T(z, t)/\partial z$ is
\begin{equation}
\widetilde{q} (x,s) = 
- q_0\ x^{2\lambda+1}  \partial \widetilde{T} (x,s)/\partial x,             \label{q}
\end{equation}
where $q_0 = \mu b z_s^{\beta-1}$. Using equation~(\ref{GenT}), we find
\begin{equation}
\widetilde{q} (x,s) = q_0 x^{\lambda+1} \left[ D_1(s) { K}_{\lambda+1}(x) - D_2(s) {
I}_{\lambda+1}(x) \right].
\label{Genq}
\end{equation}

Now let us construct the Green's function $G(z,t)$
of  equation~(\ref{a:diff1}) with the source function
\begin{equation}
Q(z, t) = \delta(z-z_\text{h}) \, \delta(t-t_\text{h}) \, H_0,
\label{delta}
\end{equation}
where $H_0$ [erg cm$^{-2}$] is the total column 
heat (per 1 cm$^2$). The source is assumed to be active at $t=t_\text{h}$ 
on a spherical shell at $z=z_\text{h}$. We are looking for the temperature $T(z,t)$ 
determined by diffusion of the generated heat at $t>t_\text{h}$ 
from the source ($z=z_\text{h}$) to small $z$ (to the stellar surface) and
to large $z$ (to the stellar interior).

Going from variable $z$ to $x$ in
equation (\ref{a:diff1}) with $Q(z, t)$ from (\ref{delta}),
then taking the Laplace transform (\ref{Laplace})
of  (\ref{a:diff1})
and using the definition (\ref{q}) in the second (transformed) term 
on the left-hand side of   (\ref{a:diff1}), 
we obtain the equation for 
$\widetilde{T} (x,s) \equiv \widetilde{G}(x,s)$,  
\begin{equation}
\partial \widetilde{q}(x,s)/\partial x
+  q_0  x^{1+2 \lambda}\ \widetilde{G}(x,s) =H_0 \delta (x-x_\text{h}) \exp(-s t_\text{h}).  
\label{dduq}
\end{equation}

In this case, equation (\ref{GenT}) has a piece-like solution,
$\widetilde{T}_{-}(x,s)$\ with coefficients $D_1^{-}(s)$ and
$D_2^{-} (s) \equiv D_{-} (s)$\ at $x < x_\text{h}$\ 
and   $\widetilde{T}_{+}(x,s)$\ with coefficients $D_1^{+} (s) \equiv D_{+} (s)$
and $D_2^{+} (s)$\  at $x > x_\text{h}$.
According to (\ref{u})  we have $x=x_\text{h}$ at $z=z_\text{h}$
{  The} two regions $z < z_\text{h}$ and $z > z_\text{h}$.
correspond to  $x < x_\text{h}$ and $x > x_\text{h}$, respectively.  

The coefficients $D_{1,2}^{\pm} (s)$ have to be determined 
from the boundary conditions. 
To proceed analytically, we introduce the following approximation that allows us to
come to the explicit solution (which is checked by comparison with
numerical simulations in Sections~\ref{s:toyvscodeAA} and \ref{s:thick}).
Instead of solving the problem in the finite interval
$z_\text{b}<z<z_\text{drip}$, we extend it to $0<z<\infty$.
Considering that 
$K_{\lambda} (x) \rightarrow 2^{\lambda -1} (\lambda -1)!\
x^{-\lambda}$, the requirement of  finite temperature at 
 $z \rightarrow 0$ (or $x \rightarrow 0$)
leads to $D_1^{-} (s) = 0$. On the other hand,  at
$ z \rightarrow \infty$, considering that   
$I_{\lambda} (x) \rightarrow e^x /\sqrt{2 \pi x}$\
 we should put $D_2^{+} = 0$.
Accordingly, the piece-like solution in the two regions becomes
\begin{subequations}
\label{DDD}  
\begin{eqnarray}&&\hspace*{-2em}
\widetilde{T}_- (x,s) = x^{-\lambda} D_-(s) J_-(x) \quad {\rm at~}x<x_\text{h}\  
(z < z_\text{h}),
\\ &&\hspace*{-2em}
\widetilde{T}_+ (x,s) = x^{-\lambda} D_+(s) J_+(x) \quad {\rm at~}x>x_\text{h}\
(z<z_\text{h}),
\hspace*{3em}
\end{eqnarray}
\end{subequations}
where $x_\text{h}=u_{h} \sqrt{s}$, $u_\text{h}$ is the same as $u$ in equation (\ref{u})
but with $z \to z_\text{h}$,
$J_-(x) \equiv I_{\lambda}(x)$, $J_+(x) \equiv K_{\lambda} (x)$. 

Integrating equation (\ref{dduq}) over an infinitesimal vicinity  
of $x=x_\text{h}$, we have
\begin{equation}
\widetilde{q}_+(x_\text{h},s) - \widetilde{q}_-(x_\text{h},s)
 = H_0 \exp(- s t_\text{h}). 
\label{q-q}
\end{equation}
In the same vicinity
we can rewrite equation (\ref{q-q}) 
using (\ref{q})
as a first order differential 
equation 
\begin{equation}
\frac{\partial \widetilde{G} (x,s)}{\partial x} = -  
{H_0  \exp(-s t_\text{h}) \over q_0 x^{2 \lambda +1}}.
\label{dGdu}
\end{equation}
Integration of (\ref{dGdu}) over the same infinitesimal
vicinity of $x_\text{h}$ gives
\begin{equation}
\widetilde{G}_+(x_\text{h},s) = \widetilde{G}_-(x_\text{h},s). 
\label{G-G}
\end{equation}
The boundary conditions  (\ref{q-q}) and (\ref{G-G})
connect solutions at
$0<x <  x_\text{h}$  and $x > x_\text{h}$. 
Combining the solutions 
(\ref{GenT}) and (\ref{Genq}), 
we have
\begin{subequations}
\begin{eqnarray}&&\hspace*{-2em}
D_+(s) { K}_{\lambda}(x_\text{h}) = D_-(s) {I}_{\lambda}(x_\text{h}) ,
\\&&\hspace*{-2em}
D_+(s) { K}_{\lambda+1}(x_\text{h}) + D_-(s) {I}_{\lambda+1} (x_\text{h})  =  
H_0 { \exp(-s t_\text{h}) \over q_0 x_\text{h}^{\lambda+1}} ,
\hspace*{3em}
\end{eqnarray}
\label{link}
\end{subequations}
which gives
\begin{equation}
D_\sigma(s)=\frac{ H_0 \exp(-s t_\text{h})}{q_0 x_\text{h}^{\lambda} } \, J_{-\sigma}(x_\text{h}), 
\label{D1D1}
\end{equation}
with $\sigma=\pm$ and $-\sigma=\mp$. Substituting  $D_\sigma(s)$ into equation~(\ref{GenT}),
we have 
\begin{equation}
\widetilde{G}_\sigma (x) = {H_0 \exp({-s t_\text{h}}) \over q_0 (x x_\text{h})^{\lambda}} 
J_{-\sigma} (x_\text{h})\ J_{\sigma} (x).        
\label{tG}  
\end{equation}

Finally, inverting the Laplace 
transform  and using the identity $q_0 (x x_\text{h})^{\lambda}= {\mu b}\  (z z_\text{h})^{(\beta-1)/2}$,
we obtain the Green's function, 
\begin{equation}
G_\sigma(z, \tau) =  \frac{H_0}{\mu b (z z_\text{h})^{(\beta-1)/2}} 
\, \mathcal{L}_{\sigma} (z, \tau), 
\label{Gztau}
\end{equation}
where
\begin{equation}
\mathcal{L}_\sigma (z, \tau)     =  
{1 \over 2\mathrm{\pi} \mathrm{i}}\ \int_{\gamma - \mathrm{i} \infty}^{\gamma +
\mathrm{i} \infty}
\dd s\ \exp({s \tau})\  
J_{-\sigma} (u_\text{h} \sqrt{s})\ J_{\sigma} (u \sqrt{s}) . 
\label{JJ}
\end{equation}
Here $\tau=t-t_\text{h} >0$, $\gamma$ is real and placed to 
the right of all singular points of the integrand 
on the imaginary $s$-plane.

To integrate in equation (\ref{JJ}) we use an integral representation
of the product  $I_{\lambda}(x) K_{\lambda}(X)$ with $X>x$
[\citealt{1953BE}, equation 7.7.6.(37)],
\begin{eqnarray}
{\cal M}
&\equiv& {1 \over 2} \int_0^\infty {\dd y \over y}\ 
\exp\left(- \frac{y}{2}- s\ \frac{u^2 + u_\text{h}^2 }{ 2 y }\right)\ 
I_{\lambda} \left({s}\  {u_\text{h}u \over y} \right)
\nonumber\\
 &=& \left\{ \begin{array}{ll}
  I_{\lambda} (u_\text{h} \sqrt{s})\ {K}_{\lambda} (u \sqrt{s}),
  & u>u_\text{h};
  \\
  I_{\lambda} (u \sqrt{s})\ {K}_{\lambda} (u_\text{h} \sqrt{s}),
  & u<u_\text{h}.
  \end{array} \right.
\label{M}         
\end{eqnarray}
We extend ${\cal M}(z, s)$ as a function of $s$
analytically along a purely imaginary axis in the $s$-plane
from $\gamma - \mathrm{i} \infty$ to $\gamma + \mathrm{i} \infty$.
Then we use in equation~(\ref{M}) 
an integral representation of the modified Bessel function
$I_{\lambda}(u_\text{h}u s/y)$ 
(e.g. \citealt{2007GR}, equation~8.431.5)
and present
${\cal M}={\cal M}(z, s)$ in the  form   
\begin{eqnarray}
{\cal M} & = &  {1 \over 2 \mathrm{\pi}} \int_0^\infty {\dd y \over y}\
\exp\left(-{y \over 2}- s\ {u^2 + u_\text{h}^2 \over 2 y}\right)
\nonumber         \\            
& \times &  \left[  \int_0^{\pi} {\rm d} x\ \cos\left(\lambda x \right)
\exp \left( {s}\, {u_\text{h}u  \over y} \,  \cos x \right)  -{\sin (\lambda \mathrm{\pi})}  \right.  
\nonumber    \\   
& \times &  \left. \int_0^{\infty} {\rm d}x\  
\exp \left( -{\lambda x} - {s}\ {u_\text{h}u\  \over y}\ \cosh x \right) \right].
\label{Iint}    
\end{eqnarray}

Using equation~(\ref{Iint}) and 
rearranging  the order of integration in equation~(\ref{JJ}), we obtain 
\begin{eqnarray}
J_{\pm} (z, \tau) & = & {1 \over 2 \mathrm{\pi}} \int_0^\infty {{\rm d} y \over
y}\ 
\exp \left(- \frac{y}{2} \right)  
\left[
\int_0^{\pi} {\rm d} x\ \cos \left(\lambda x  \right)\ \Delta_1 \right.
\nonumber \\
& - & \left.  \sin (\lambda \mathrm{\pi}) \int_0^{\infty} {\rm d} x\  \exp \left(- {\lambda x
} \right)
\Delta_2 \right],     
\label{Jpm}
\end{eqnarray}
where 
\begin{equation}
\Delta_1  =   {1 \over 2\mathrm{\pi} \mathrm{i}} \int_{\gamma -
\mathrm{i}\infty}^{\gamma + \mathrm{i} \infty}
{\rm d} s \  
\exp\left[ s \left( \tau - {u^2 + u_\text{h}^2 \over 2y}  + 
{u_\text{h}u  \over y} \cos x \right) \right],
\nonumber   
\end{equation}
and $\Delta_2$ is obtained from $\Delta_1$ by replacing $\cos x \to -\cosh x$.

In the integral over $s$ we introduce a real variable $\omega=\mathrm{i}s$. 
This results in real values of $J_{\pm} (z, \tau)$ and justifies 
employing  the integral representation
for $I_{\lambda} (x)K_{\lambda} (y)$ in equation~(\ref{M}).    
Then $\Delta_1$ is expressed via the Dirac delta-function, 
\begin{equation}
\Delta_1  =  {y \over \tau} \delta \left( y - {(u - u_\text{h})^2 \over 2 \tau} -
{u_\text{h}u  \over \tau} (1- \cos x) \right),
\label{Delta1-delta} 
\end{equation}
and a similar expression is valid for for $\Delta_2$, with $\cos x \to -\cosh x$.

Furthermore, after trivial integrations over $y$ in equation~(\ref{Jpm}) 
we are left with the integration over $x$  
which is carried out  using the same integral representation as in 
equation~(\ref{Iint}),
\begin{equation}
J_{\pm}={1 \over 2\tau} \exp\left[-{(u-u_\text{h})^2 \over 4 \tau}\right] 
\exp\left(-{u_\text{h}u  \over 2 \tau} \right) 
{ I}_{\lambda} \left({u_\text{h}u  \over 2 \tau} \right).
\label{intdx}
\end{equation}

Then employing  equations (\ref{Gztau}), (\ref{intdx}) and (\ref{u}),
we come to the final expression for the Green's function
\begin{equation}
G(z,\tau)  =    {H_0 \over 2 \mu b \tau (z z_\text{h})^{(\beta-1)/2}} 
\exp \left( -\frac{u^2 + u_\text{h}^2}{4 \tau } \right)
{ I}_{\lambda} \left({u_\text{h}u  \over 2 \tau} \right).
\label{a:finalG}
\end{equation}
A similar Green's function was derived by \citet{ec89} 
(although without proper normalization).

In the bulk of this paper we have used the toy model with
$\alpha=3$, $\beta=2$, $\mu=3/2$ and $\lambda=1/3$. Then
\begin{equation}
G(z,\tau)  =    {H_0 \over 3 b \tau \sqrt{z z_\text{h}}} 
\exp \left( -\frac{u^2+u_\text{h}^2 }{ 4 \tau } \right)
{ I}_{1\over 3} \left({u_\text{h}u  \over 2 \tau} \right),
\label{finalG}
\end{equation}
which is essentially the same as (\ref{e:Green}).

A more general solution (\ref{a:finalG}) can be used to describe heat diffusion 
in some local stellar layers where the heat capacity $C$ and thermal conductivity $\kappa$
are independent of temperature but depend on density. One can also obtain similar
analytic solutions if (in addition to the density dependence) $C$ and $\kappa$ 
are power-law functions of temperature with the same power index.

\label{lastpage}

\end{document}